\documentclass[fleqn,usenatbib]{mnras}

\usepackage[T1]{fontenc}
\usepackage{ae,aecompl}

\usepackage{graphicx}	
\usepackage{amsmath}	
\usepackage{amssymb}

\usepackage{txfonts}

\newcommand{\btxt}[1]{{#1}}	

\title[Ancient eruptions of $\eta$ Car]{Ancient eruptions of $\eta$ Carinae: A tale written in proper motions}

\author[M. M. Kiminki et al.]{
Megan M. Kiminki,$^{1}$\thanks{E-mail: \href{mailto:mbagley@email.arizona.edu}{mbagley@email.arizona.edu}}
Megan Reiter$^{2}$
and Nathan Smith$^{1}$
\\
$^{1}$Steward Observatory, University of Arizona, 933 N.
  Cherry Avenue, Tucson, AZ 85721, USA\\
$^{2}$Department of Astronomy, University of Michigan, 311 West Hall,
  1085 S. University Avenue, Ann Arbor, MI 48109, USA }

\date{Accepted 2016 August 9. Received 2016 August 8; in original form 2016 June 6.}

\pubyear{2016}

\begin{document}
\label{firstpage}
\pagerange{\pageref{firstpage}--\pageref{lastpage}}
\maketitle

%%%%%%%%%%%%%%%%%%%%%%%%%%%%%%%%%%%%%%%%%%%%%%%%%%%%%%%%%%%%%%%%%%%%%%%
\begin{abstract}
We analyze eight epochs of \emph{Hubble Space Telescope}
H$\alpha$+[\ion{N}{ii}] imaging of $\eta$ Carinae's outer ejecta.
Proper motions of nearly 800 knots reveal that the
\btxt{detected} ejecta are divided into three \btxt{apparent} age
groups, \btxt{dating to around 1250 A.D., to around 1550 A.D., and to
  during or shortly before the Great Eruption of the 1840s.}  Ejecta
from these groups reside in different locations and provide a firm
constraint that $\eta$ Car experienced multiple major eruptions prior
to the 19\textsuperscript{th} century.  The 1250 and 1550 events did
not share the same axisymmetry as the Homunculus; the 1250 event was
particularly asymmetric, even one-sided.  In addition, the ejecta in
the S ridge, which have been associated with the Great Eruption,
appear to predate the ejection of the Homunculus by several decades.
\btxt{We detect essentially ballistic expansion across multiple
  epochs.}  We find no evidence for large-scale deceleration of the
\btxt{observed} knots that could power the soft X-ray shell by plowing
into surrounding material, suggesting that the observed X-rays arise
instead from fast, rarefied ejecta from the 1840s overtaking the older
dense knots.  \btxt{Early deceleration and subsequent coasting cannot
  explain the origin of the older outer ejecta---significant episodic
  mass loss prior to the 19\textsuperscript{th} century is required.}
The timescale and geometry of the past eruptions provide important
constraints for any theoretical physical mechanisms driving $\eta$
Car's behavior.  Non-repeating mechanisms such as the merger of a
close binary in a triple system would require additional complexities
to explain the observations.
\end{abstract}


%%%%%%%%%%%%%%%%%%%%%%%%%%%%%%%%%%%%%%%%%%%%%%%%%%%%%%%%%%%%%%%%%%%%%%%
\section{Introduction}
\label{sec:intro}

One of the most remarkable stars in our galaxy, $\eta$ Carinae has
been puzzling astronomers for over 150 years.  In the mid-nineteenth
century, it became increasingly variable, then peaked temporarily as
the second brightest star in the sky
\citep{innes1903,davidsonhumphreys1997,frew2004,smithfrew2011} before
slowly fading over more than a decade.  During this Great Eruption,
$\eta$ Car ejected an estimated 10--15 M$_{\sun}$ into the well-known
bipolar Homunculus nebula \citep{smith2003}.  A second, Lesser
Eruption followed in 1890 \citep{innes1903,humphreys1999,frew2004},
but only ejected $\sim0.1$ M$_{\sun}$ \btxt{\citep{ishibashi2003,smith2005a}}.

$\eta$ Car belongs to a class of stars known as luminous blue
variables \citep[LBVs,][]{humphreysdavidson1994}, very massive,
unstable, post-main-sequence stars characterized by luminous mass-loss
events.  Even among LBVs, $\eta$ Car is unusual and its parameters are
extreme.  In its current quiescent state, $\eta$ Car is substantially
more luminous than most other known LBVs
\citep[e.g.,][]{vangenderen2001,smithtombleson2015}.  It is one of
only two Galactic LBVs that has been observed in a giant eruption.
The other is P Cygni, whose largest eruption involved significantly
less energy and mass loss, similar to $\eta$ Car's Lesser Eruption
\citep{smithhartigan2006}.  Moreover, $\eta$ Car has a massive binary
companion in an eccentric 5.5-year orbit
\citep{damineli1996,damineli1997,damineli2000,corcoran2001,whitelock2004},
and is located in the rich cluster Trumpler 16, home to dozens of
O-type stars \citep{smith2006a}.  In contrast, most LBVs are
relatively isolated and lack O-type neighbors
\citep{smithtombleson2015}.

The mechanism of $\eta$ Car's Great Eruption---which released roughly
$10^{50}$ ergs of kinetic energy \citep{smith2003,smith2008}---remains
a mystery.  Many theories treat it as part of single-star evolution,
invoking super-Eddington radiation-driven winds
\citep{davidson1971,maeder1983,dejager1984,lamersfitzpatrick1988,stotherschin1993,glatzelkiriakidis1993,glatzel1994,humphreysdavidson1994,shaviv2000,owocki2004}.
However, the source of the increased bolometric luminosity in these
scenarios is unclear.  Alternatively, the eccentric orbit of $\eta$
Car's companion has been taken to imply that the Great Eruption was
influenced by periastron interactions between the two binary
components. Based on nineteenth-century observers' estimates of the
primary star's color and brightness, its radius must have been much
larger than at present, large enough that its companion would
significantly interact or even physically collide
\citep{iben1999,smith2011}.  \btxt{A speculative idea is that} the
collision mixed fresh nuclear fuel to greater depths, causing a sudden
burst of increased nuclear burning \citep{smith2011}.  It has also
been proposed that periastron tidal interactions spun up the primary
to unstable rates, leading to a burst of mass loss
\citep{cassinelli1999}, or that the Great Eruption was fueled by
accretion from the primary onto its companion
\citep{soker2007,kashisoker2010}.  Still other theories postulate a
hierarchical triple system, in which the close inner pair either
merged
\citep{gallagher1989,iben1999,morrispodsiadlowski2009,podsiadlowski2010,portegieszwartvandenheuvel2016}
or underwent a dynamical exchange with the outer companion
\citep{liviopringle1998}.

\begin{figure}
  \includegraphics[width=\columnwidth]{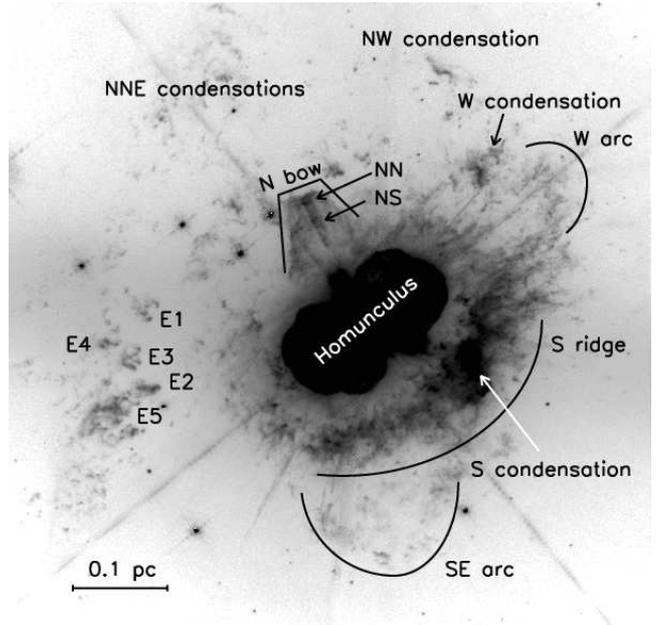}
  \caption{\emph{HST} WFPC2 image of $\eta$ Car in the F658N filter,
    which captures intrinsic and scattered [\ion{N}{ii}] $\lambda$6584
    emission along with redshifted H$\alpha$ emission
    \citep{morse1999,morse2001}.  Prominent features are labeled
    according to the convention of \citet{walborn1976} and
    \citet{weis2012}.}
  \label{fig:labels}
\end{figure}

Models of the driving cause of the Great Eruption must also \btxt{incorporate}
the outer ejecta, a collection of irregular condensations found out to
nearly half a parsec outside the Homunculus
\citep{thackeray1950,walborn1976,meaburn1996a,smithmorse2004,weis2012}.
These outer ejecta (Figure \ref{fig:labels}) are highly nitrogen-rich,
suggesting a substantial degree of CNO processing
\citep{davidson1986,smithmorse2004}.  They contain a minimum mass of
2--4 M$_{\sun}$ \citep{weis2012}, with dust observations suggesting a
much larger total mass \citep{gomez2010}.  The various models for the
Great Eruption produce different explanations for the outer ejecta.
The merger model of \citet{portegieszwartvandenheuvel2016}, for
instance, predicts that the outer ejecta were formed after the
formation of the Homunculus.

The proper motions of the outer ejecta provide concrete constraints on
$\eta$ Car's mass-loss history.  The bright S condensation and the
``jet''-shaped N bow (see Figure \ref{fig:labels}) have motions
consistent with having been ejected during the Great Eruption
\citep{walborn1978,ebbets1993,currie1996,morse2001}.  Some results
have suggested, however, that the extended S ridge is up to one
hundred years older \citep{walborn1978,morse2001}.  The age of the E
condensations is even less clear: \citet{walborn1978} found transverse
velocities of 300--400 km s$^{-1}$, indicating ejection dates in the
mid-1400s, but \citet{walbornblanco1988} determined ten years later
that the same features had slowed dramatically, suggesting they were
from the Great Eruption after all. While the motions of the
outer condensations have hinted at prior mass-loss events, a single
ejection date around the time of the Great Eruption could not be ruled
out.

%----------------------------------------------------------------------
%----------------------------------------------------------------------

In this paper, we measure the proper motions of $\eta$ Car's outer
ejecta to unprecedented accuracy, using 16 different baselines over 21
years of \emph{Hubble Space Telescope} (\emph{HST}) data.  The depth
and resolution of the \emph{HST} images allow us to re-evaluate the
origins of the N, E, and S features, and, for the first time, measure
motions of the fainter NNE, NW, and SE condensations.  We find no
evidence of widespread deceleration, and show that \btxt{while some
  of} the outer ejecta come from the Great Eruption (or the lead-up to
it), \btxt{many features require} at least one prior mass-loss event
centuries earlier.  Our data, image registration, and approach to
measuring proper motions are described in Section \ref{sec:obs}; the
results are presented in Section \ref{sec:results}.  We discuss the
implications of our results on models of $\eta$ Car in Section
\ref{sec:disc} and conclude with a summary in Section \ref{sec:conc}.

%%%%%%%%%%%%%%%%%%%%%%%%%%%%%%%%%%%%%%%%%%%%%%%%%%%%%%%%%%%%%%%%%%%%%%%
\section{Observations and Analysis}
\label{sec:obs}

\subsection{\emph{HST} ACS}
\label{subsec:acs}

%----------------------------------------------------------------------
\begin{table*}
  \centering
  \caption{\emph{HST} data log.}
  \label{tab:obstab}
  \begin{tabular}{llllcr}
    \hline
    Instrument & Camera & Date &  Filter & Exp. Time (s) & Program ID \\
    \hline
    WFPC2 & WF3 & 1993 Dec 31  & F658N & 2 $\times$ 200 & 5188 \\
    WFPC2 & WF3 & 1997 Jul 11  & F658N & 2 $\times$ 200 & 7253 \\
    WFPC2 & WF3 & 1999 Jun 12  & F658N & 2 $\times$ 200 & 8178 \\
    WFPC2 & WF3 & 2001 Jun 4   & F658N & 2 $\times$ 200 & 9226 \\
    WFPC2 & WF3 & 2003 Aug 8   & F658N & 2 $\times$ 100 & 9775 \\
    \vspace{3pt}
    WFPC2 & WF3 & 2008 Sept 6   & F658N & 2 $\times$ 100 & 11500 \\
    ACS & WFC & 2005 Jul 18 & F658N$^{\mathrm{a}}$  & 2 $\times$ 500 & 10241 \\
    ACS & WFC & 2014 Aug 4  & F658N$^{\mathrm{a}}$  & 2 $\times$ 520 & 13390 \\
    \hline
  \end{tabular}
  \\
  \medskip
  \footnotesize
  \begin{tabular}{l}
    \btxt{$^{\mathrm{a}}$Note that the bandpass of the ACS 658N filter is
    different from that of the WFPC2 filter of the same name (see
    discussion in Section \ref{subsec:wfpc2}).} \\
    \end{tabular}
\end{table*}
%----------------------------------------------------------------------

We obtained new H$\alpha$+[\ion{N}{ii}] images of $\eta$ Car and the
surrounding Tr 16 cluster on 2014 Aug 4, using the Wide Field Channel
(WFC) of the Advanced Camera for Surveys (ACS) on \emph{HST} (program
ID 13390; see Table \ref{tab:obstab}).  These observations were
designed to replicate our observations of 2005 Jul 18
\citep{smith2010a} as closely as possible, in pointing and position
angle, in order to minimize position-dependent uncertainties when
measuring proper motions.  The Homunculus and central star are heavily
saturated in these exposures, which were each 500--520 s long.  In both
epochs, observations were made as a series of $205\times400$
arcsec$^{2}$ ``footprints,'' each consisting of three pairs of {\tt
  CR-SPLIT} dithers in a linear offset pattern designed to fill in the
ACS chip gaps.  $\eta$ Car and its outer ejecta were observed in a
single footprint; the rest make up a mosiac of Tr 16 and neighboring
Tr 14 \citep[see][]{smith2010a}. All data were processed by the
standard \emph{HST} ACS pipeline, which does bias subtraction and
flat-fielding and corrects for charge transfer efficiency (CTE) to
produce an image with the designation ``{\tt flc}.''  The pipeline also
produces a ``{\tt drc}'' image that has been additionally corrected
for geometric distortion and dither-combined via AstroDrizzle.

%----------------------------------------------------------------------
\begin{figure*}
  \centering
  \frame{\includegraphics[width=0.495\textwidth, trim=-3mm -2mm 3mm -2mm,clip]{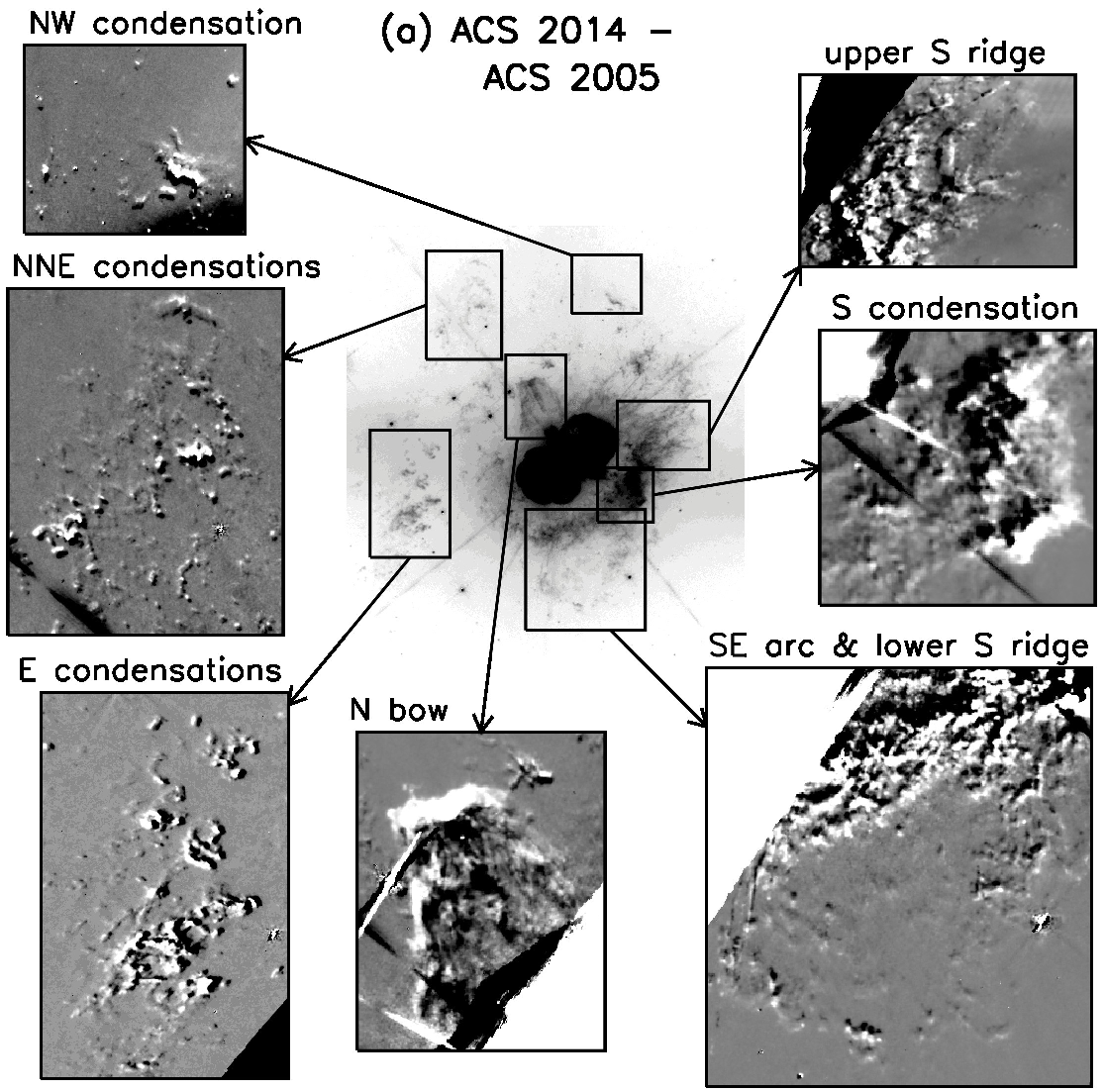}}
  \frame{\includegraphics[width=0.495\textwidth, trim=-3mm -2mm 3mm -2mm,clip]{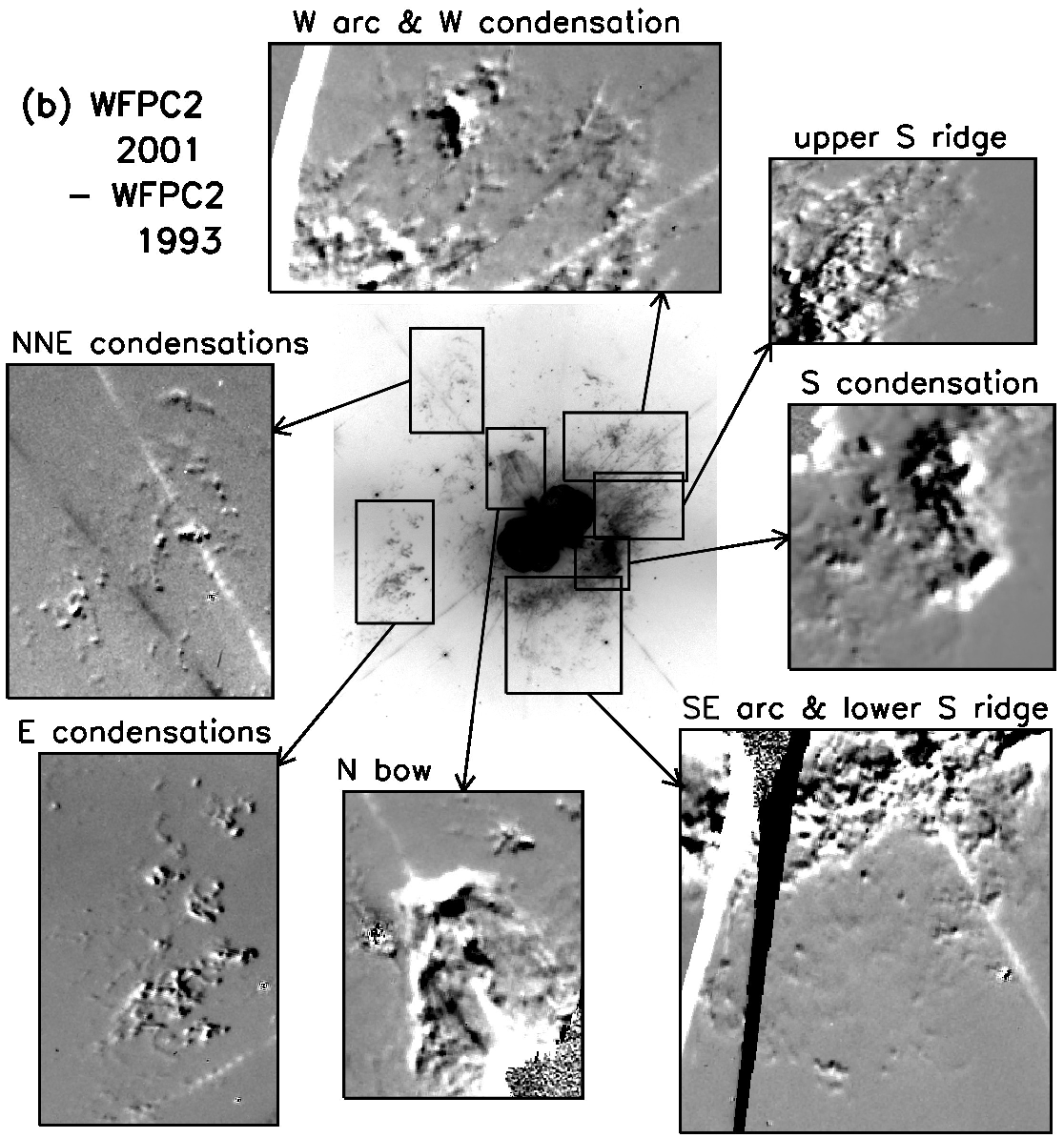}}
  \caption{(a) Difference images showing the outward motion of $\eta$
    Car's ejecta over the 9-year baseline between ACS images.  Dark
    patches are where the various condensations were in 2005; bright
    patches are places into which that material had moved by 2014.
    The central image is the same one shown in Figure \ref{fig:labels}
    and is included to orient the reader.  (b) Same as (a) but for the
    7-year baseline between the WFPC2 1993 and 2001 images.  The W and
    NW condensations are not shown in the ACS (left) and WFPC2 (right)
    panels, respectively, because they are contaminated by bleeding
    from the saturated Homunculus in the relevant epochs.  Animated
    GIFs showing the motion of each feature across all 8 epochs are
    available in the online supplementary material.
  \label{fig:diffimages}}
\end{figure*}

%----------------------------------------------------------------------

We transformed both epochs of ACS data to a common distortion-free
reference frame via a modified version of the method used in
\citet{anderson2008a,anderson2008b}, \citet{andersonvandermarel2010},
and \citet[][see also \citealt{reiter2015a,reiter2015b}]{sohn2012}.
An initial reference frame, aligned with the $y$ axis pointing north
and with a pixel scale of 50 mas pixel$^{-1}$, was constructed using
the astrometric solutions in the headers of the {\tt drc} images to
match stars that appear in overlapping images.  However, the
resampling performed by AstroDrizzle makes the {\tt drc} images
unsuitable for direct high-accuracy positional measurements.  Instead,
we performed point spread function (PSF) fitting on the the undrizzled
{\tt flc} images using the program {\tt img2xym\_WFC.09x10}
\citep{andersonking2006}, which uses a library of spatially dependent
effective PSFs.  We then applied the geometric distortion corrections
of \citet{anderson2006} to the measured stellar positions.

Next, we iteratively mapped the stellar positions from the {\tt flc}
images to the reference frame by:
\begin{enumerate}
  \item identifying the high signal-to-noise stars (typically several
    hundred per image) in common between each {\tt flc} image and the
    reference frame;
  \item computing the six-dimensional linear transformation from the
    distortion-corrected positions of the {\tt flc} images to the
    reference frame; and
  \item replacing each existing reference-frame stellar position with
    the average of that star's transformed, distortion-corrected {\tt
      flc} positions.
\end{enumerate}
After three iterations, the internal accuracy of the reference frame
was $<0.02$ pixels (1 mas, or 1.2 km s$^{-1}$ over this 9-year
baseline at the distance of $\eta$ Car) across a single footprint.

Using the final linear transformations determined by this process, we
resampled the {\tt flc} images from each epoch into the reference
frame using the algorithm described in \citet{anderson2008a}, scaling
each image to a total exposure time of 500 s.  The end result was a
single stacked image at each epoch for each observed footprint.  With
both epochs of data on the same reference frame, the positions of
ejecta features in the stacked images could be directly compared.
Figure \ref{fig:diffimages} presents difference images (stacked image
from 2014 minus stacked image from 2005) for named regions of
interest in $\eta$ Car's outer ejecta.  In these images, material has
moved from the dark patches into the corresponding bright areas.
It is immediately qualitatively apparent that some features have
greater transverse velocities than others; the N bow, for instance, is
rapidly overtaking the small feature to its immediate northwest.
Notably, there are no pronounced changes in condensation morphology or
brightness from 2005 to 2014.

\begin{figure}
  \centering
  \includegraphics[width=0.9\columnwidth]{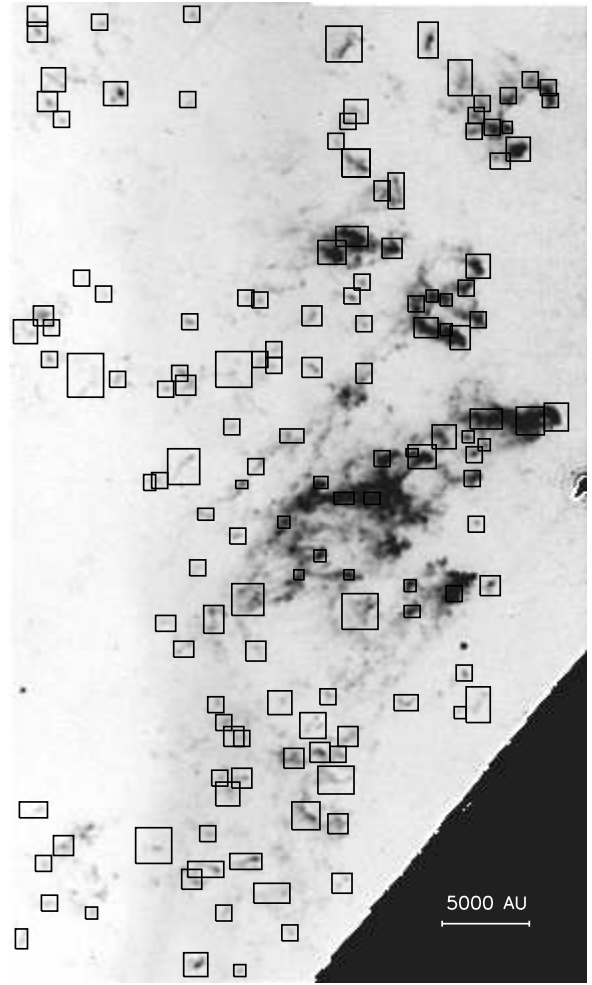}
  \caption{Close-up of the region around the E condensations in the
    2005 ACS F658N image, showing the boxes used to measure proper
    motions via modified cross-correlation.  The
    saturated strip in the lower right is bleeding from the saturated
    Homunculus.
    \label{fig:boxes}}
\end{figure}

The proper motions of well-defined knots in the ejecta around $\eta$
Car were measured in the stacked images.  ``Well-defined'' features
are those that are (1) not contaminated by or confused with the
saturated Homunculus or diffraction spikes in either epoch and (2)
sufficiently isolated for a defined knot or feature to be identified.
A total of 620 such features were identified in the ACS images.  To
measure the proper motion of each, we subtracted a median-filtered
image (filtered using a kernal size of 12 pixels) in order to suppress
the local diffuse H$\alpha$ background.  We then extracted the knot
using a box size optimized for that feature; Figure \ref{fig:boxes}
illustrates the boxes used for the features in and around the E
condensations. We used our implementation of the modified
cross-correlation technique developed by \citet{currie1996},
\citet{hartigan2001}, and \citet[][see also
  \citealt{reitersmith2014,reiter2015a,reiter2015b}]{morse2001} to
determine the difference in position between the two epochs.  In
brief, the box containing the feature of interest was compared to the
second image and an array was generated containing the total of the
square of the difference between two images for each shift. The
minimum of this array corresponds to the shift that best matches the
two images.  Angular displacements were converted to km s$^{-1}$
assuming a distance of 2.3 kpc \citep{smith2006b}.  To estimate the
uncertainty on the offset, we repeated the procedure using a variety
of box sizes.  The median proper motion uncertainty for features
measured in the ACS images is 2.7 km s$^{-1}$.

%----------------------------------------------------------------------
%----------------------------------------------------------------------

%----------------------------------------------------------------------
\begin{table}
  \centering
  \caption{Baselines used for proper motion measurements.}
  \label{tab:bases}
  \begin{tabular}{lrrrr}
    \hline
    Baseline &  Instrument & $\Delta$t & \# Features &  Median $\sigma$  \\
     & & (years) & & (km s$^{-1}$) \\
    \hline
     1993--1997 & WFPC2 &  3.524068 & 694 &  8.5 \\
     1993--1999 & WFPC2 &  5.443602 & 704 &  7.0 \\
     1993--2001 & WFPC2 &  7.422813 & 712 &  5.4 \\
     1993--2003 & WFPC2 &  9.599420 & 714 &  4.8 \\
     1993--2008 & WFPC2 & 14.680967 & 698 &  3.9 \\
     1997--1999 & WFPC2 &  1.919534 & 703 & 14.3 \\
     1997--2001 & WFPC2 &  3.898745 & 742 &  7.7 \\
     1997--2003 & WFPC2 &  6.075352 & 713 &  6.3 \\
     1997--2008 & WFPC2 & 11.156898 & 686 &  4.6 \\
     1999--2001 & WFPC2 &  1.979211 & 716 & 12.1 \\
     1999--2003 & WFPC2 &  4.155818 & 684 &  8.0 \\
     1999--2008 & WFPC2 &  9.237365 & 679 &  4.6 \\
     2001--2003 & WFPC2 &  2.176607 & 633 & 13.0 \\
     2001--2008 & WFPC2 &  7.258153 & 666 &  5.5 \\
     \vspace{3pt}                           
     2003--2008 & WFPC2 &  5.081546 & 648 &  7.6 \\
     2005--2014 & ACS   &  9.046256 & 620 &  2.7 \\
     \hline
  \end{tabular}
\end{table}
%----------------------------------------------------------------------

\subsection{\emph{HST} WFPC2}
\label{subsec:wfpc2}

To supplement our proper motion measurements from the 2005--2014 ACS
baseline, we searched the \emph{HST} archive for additional deep
images of $\eta$ Car taken with an F658N filter.  Six epochs of
observations from the Wide-Field Planetary Camera 2 (WFPC2) met our
requirements: the images had to be deep enough for the outer ejecta to
be visible, and $\eta$ Car had to be centered on one of the Wide Field
chips (the field of view of the Planetary Camera chip is too small).
The dates and program IDs for these data are given in Table
\ref{tab:obstab}.  We retrieved the longest exposures available at
each epoch: a pair of 200-s exposures from each of 1993, 1997, 1999,
and 2001, and a pair of 100-s exposures from each of 2003 and 2008.
$\eta$ Car itself and much of the Homunculus are saturated in these
images, although the bleeding is much less extensive than in the ACS
images, which have both longer exposure times and a wider filter
bandpass (see below).  All of the data had been reprocessed by the
final version of the \emph{HST} WFPC2 pipeline, which produces images
designated ``{\tt c0m}'' that have been bias-subtracted,
dark-corrected, and flat-fielded.

Note that there \btxt{are important differences} between the WFPC2
F658N filter and the ACS filter of the same name.  The former covers a
narrow wavelength range (FWHM 38 \AA) roughly centered on
[\ion{N}{ii}]$\lambda6584$, capturing emission from that line with
Doppler shifts from approximately -615 to +1130 km s$^{-1}$.  The
latter is broader (FWHM 72 \AA), capturing [\ion{N}{ii}] emission with
radial velocities up to $\pm$1600 km s$^{-1}$ as well as significant
H$\alpha$ emission.  Although most of the same ejecta features seen in
the ACS images are identifiable in the WFPC2 images, there is no way
to guarantee that the knot shapes are unaffected by the differences in
filter bandpass.  \btxt{Consequently, we do not measure proper motions
  directly between WFPC2 and ACS images.  We use only WFPC2--WFPC2 and
  ACS--ACS baselines.}  In addition, the fastest known ejecta
\btxt{are Doppler-shifted out of one or both filters and are not
  detected.  The WFPC2 F658N filter would not, for instance, pick up
  the -875 km s$^{-1}$ blueshifted emission just outside the southeast
  lobe of the Homunculus \citep{currie2002}, even if the Homunculus
  were not saturated in our images.  Both filters miss the -3200 km
  s$^{-1}$ blueshifted ejecta detected further out by
  \citet{smith2008}.  Our images thus exclude the fastest-moving
  material from the Great Eruption and therefore do not exclude the
  possibility of additional recent mass-loss episodes.}  We note,
however, that since few of the ejecta seen in the ACS images drop out
in the WFPC2 images (and those that do are relatively faint and not
detected in the shortest exposures), few to none of the dense knots
measured in our images have radial velocities of $\pm$1000--1500 km
s$^{-1}$.

We took the same approach to aligning and stacking the WFPC2 data as
we did with the ACS images.  We used a modified version of the {\tt
  img2xym} program \citep{andersonking2006} to measure stellar
positions in the {\tt c0m} images, employing a library of
spatially dependent effective WFPC2 PSFs from
\citet{andersonking2000}.  The measured positions were then corrected
for the 34th-row anomaly \citep{andersonking1999} and for geometric
distortion \citep{andersonking2000}.  We identified 6--10 isolated,
low-proper-motion stars in common between each WFPC2 image and the
ACS-based reference frame, and used those stars to derive a
six-parameter linear transformation between each distortion-corrected
WFPC2 frame and the reference frame.  Unlike with the ACS data, we did
not iterate on the process: we took the ACS-based reference frame as
final.  The internal accuracy of the alignment and stacking procedure
is $<0.06$ reference-frame pixels (3 mas) for the first four epochs.
The shorter, lower signal-to-noise 2003 and 2008 exposures had greater
positional uncertainties of up to 0.2 reference frame pixels (10 mas).

Using the calculated transformations, the WFPC2 {\tt c0m} images were
resampled and stacked into a single image at each epoch, aligned with
the ACS-based reference frame.  The {\tt c0m} images, which have a
plate scale of 99.6 mas pixel$^{-1}$, were oversampled during stacking
so that the output image matched the 50-mas pixels of the ACS images.
An example of the results is shown in the right panel of Figure
\ref{fig:diffimages}, which shows difference images for the major
named regions of $\eta$ Car's outer ejecta for the 1993--2001
baseline.  Although the resolution is somewhat lower compared to the
ACS images, much detail is still apparent.  Animated GIFs of each
region, made from all eight epochs of aligned data, are available in
the online supplementary material.

The proper motions of the features identified in the ACS images were
measured in the WFPC2 images, using the modified cross-correlation
technique described in Section \ref{subsec:acs}, for every possible
baseline among the six epochs (a total of 15 baselines).  Some
features were contaminated by diffraction spikes or saturation bleeds
in some epochs; proper motions for these features were measured using
only the epochs in which they were not contaminated.  The baselines,
number of features measured, and median proper motion uncertainties
for each baseline are given in Table \ref{tab:bases}.  The median
proper motion uncertainties range from 3.9 to 14.3 km s$^{-1}$, and
the magnitude of the uncertainty is highly negatively correlated with
the length of the baseline (i.e., shorter baselines produce greater
uncertainties). In the areas blocked by saturation bleeding in the ACS
images, we used the 1993--2001 baseline (the longest baseline with
200-s exposures) to identify additional features which were then
measured across the other WFPC2 baselines where possible.  A small
number of features in a gap that was saturated in 2001 are identified
in the 1993--2003 baseline instead, then measured across other
baselines where possible.  In total, we measure the proper motions of
792 individual features in the ejecta of $\eta$ Car.  Each feature is
measured in 1--16 baselines (including the 2005--2014 ACS baseline);
on average, a feature is measured in 14 baselines.

\subsection{Position of the central source}
\label{subsec:etacen}

Owing to the heavy saturation of the Homunculus, we were unable to
measure the position of $\eta$ Car itself in any of our stacked
images.  Instead, we measure the centroid of the central star in an
F631N WFPC2 Planetary Camera image from \citet{morse1998}.  The
centroid positions of five surrounding stars in this image were used
to derive a linear transformation to our reference frame.  We estimate
that this transformation is accurate to $\sim0.2$ reference-frame
pixels.  Added in quadrature with the $\pm$0.5-pixel uncertainties in
absolute knot positions, this means that the overall uncertainty in a
knot's distance from $\eta$ Car is $\sim$0.55 pixels (27 mas, or 60
A.U. at the distance of $\eta$ Car).  For a knot moving at 300 km
s$^{-1}$, this translates to an uncertainty in its age of $\pm$0.9
years.

%%%%%%%%%%%%%%%%%%%%%%%%%%%%%%%%%%%%%%%%%%%%%%%%%%%%%%%%%%%%%%%%%%%%%%%
\section{Results}
\label{sec:results}

\subsection{\emph{HST} proper motions over two decades}

%----------------------------------------------------------------------
\begin{figure*}
  \centering
  \includegraphics[width=0.8\linewidth]{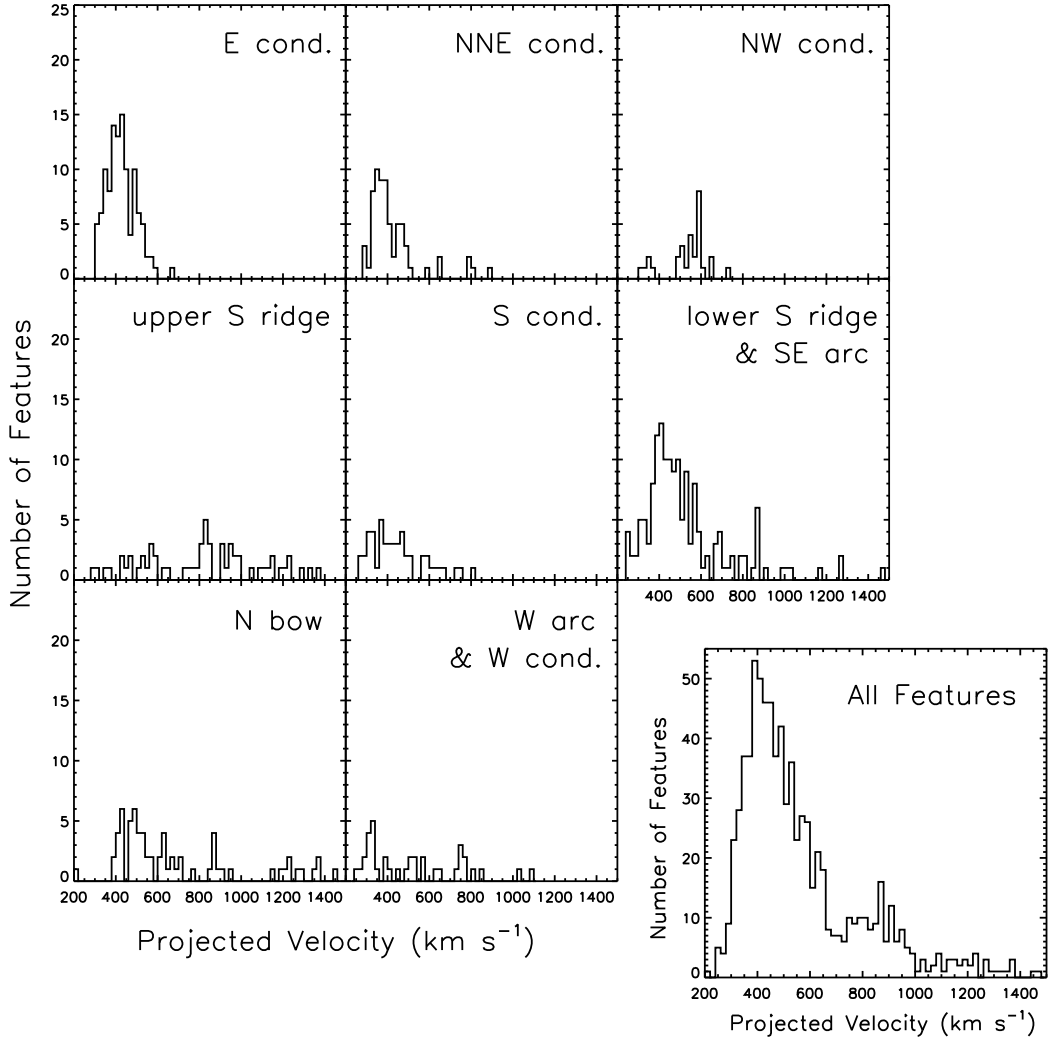}
  \caption{Histograms of the weighted mean proper motions of the outer
    ejecta, assuming a distance of 2.3 kpc.  The regions are as shown
    in Figure \ref{fig:diffimages}; note that some of the regions
    slightly overlap.  The bottom right panel shows the velocity
    histogram for all 792 measured features.
  \label{fig:velhist}}
\end{figure*}
%----------------------------------------------------------------------

Figure \ref{fig:velhist} presents our results in aggregate: a
histogram of the weighted mean proper motions of all measured
features, along with proper motion histograms for the ejecta in each
of the regions highlighted in Figure \ref{fig:diffimages}.  The
transverse velocities of $\eta$ Car's outer ejecta range from 219 km
s$^{-1}$ to 1462 km s$^{-1}$, with a broad peak at 300--600 km
s$^{-1}$ and a secondary peak at 750--900 km s$^{-1}$.  The highest
proper motions are found in the N bow, the S ridge, and the parts of
the W arc that overlap with the upper S ridge.  The presence and
location of speeds upward of 1000 km s$^{-1}$ agrees with past
observations of the S ridge and N bow
\citep{walborn1978,walbornblanco1988,ebbets1993,morse2001}.  The mix
of speeds seen in the lower S ridge is also consistent with the
results of \citet{morse2001}.

Our results for the E condensations (no high-proper-motion features;
mean proper motion $\sim400$ km s$^{-1}$) agree solidly with the
results of \citet{walborn1978}, who used photographic plate images
from 1950 to 1975.  However, the observed motion in our \emph{HST}
images disagrees strongly with the deceleration hypothesis of
\citet{walbornblanco1988}, who added data from 1985 and found that the
E condensations had dramatically slowed over 10 years.  The
\citet{walbornblanco1988} deceleration hypothesis predicts that the E
features would have reached zero velocity by the end of the twentieth
century, a hypothesis that our data definitively \btxt{exclude}.

\begin{figure}
  \centering
  \includegraphics[width=\columnwidth, trim=5mm 0 0 0, clip]{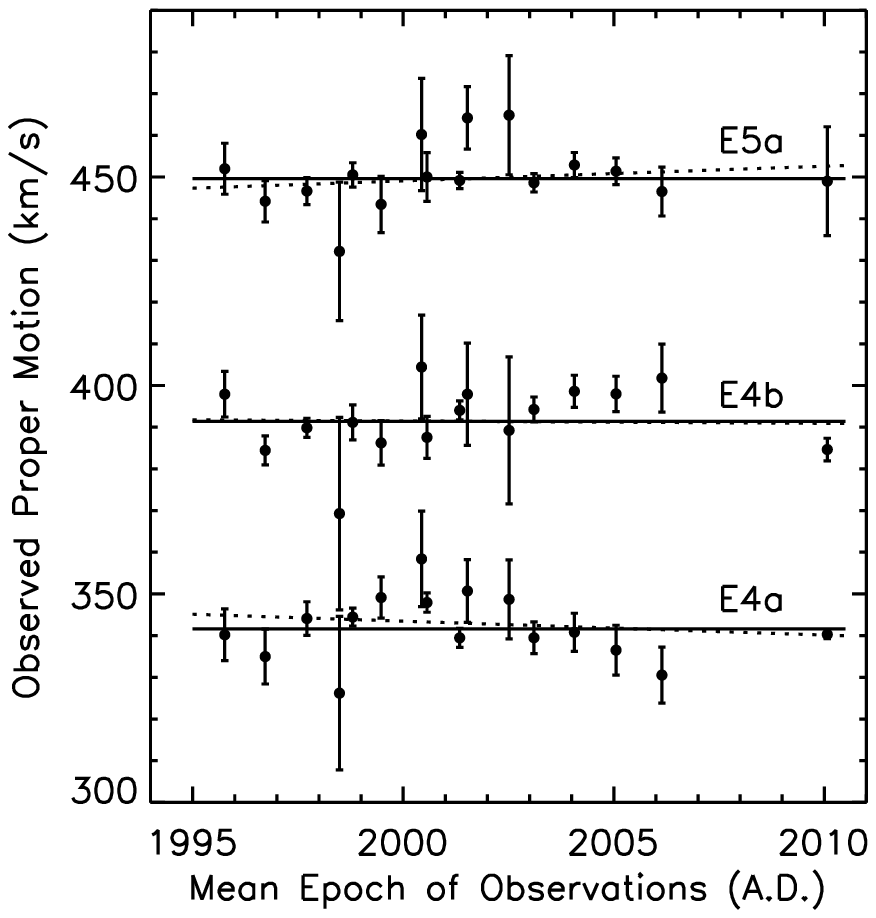}
  \caption{Proper motions measured over all possible baselines for
    three representative features among the E condensations.  The
    points around 2010 are from the ACS 2005--2014 baseline, while the
    other measurements come from the various WFPC2--WFPC2 baselines.
    The largest uncertainties typically occur over the shortest
    baselines, e.g., 1997--1999.  The solid lines are the weighted means
    of the measurements, and the dotted lines are least-squares fits to
    proper motion versus time for each ejecta feature.
  \label{fig:threeE}}
\end{figure}

With a total of 16 different baselines from two instruments, we can
approach the question of acceleration/deceleration with unprecedented
detail.  Figure \ref{fig:threeE} plots the measured proper motions
over time for three features representative of the E condensations and
of the ejecta overall.  The weighted mean transverse velocity of each
feature is overplotted, along with a least-squares fit to the data.
Although there is some variation among the proper motions, there are
no overall trends with time.  In a $\chi^2$ test, the measured motions
for all three features shown in Figure \ref{fig:threeE} are consistent
with having been drawn from a distribution with constant transverse
velocity.

%----------------------------------------------------------------------
%----------------------------------------------------------------------

This level of variance and (lack of) acceleration is found across all
our data.  Out of the 792 features for which proper motions were
measured, only 50 (6\%) have data that are inconsistent at the
$p<0.05$ level with being drawn from a constant velocity distribution.
These 50 features occupy no special region of physical space or
velocity space.  The slopes of the least-squares fits to velocity
versus time for these 50 features also fall inside the range of slopes
found for the rest of the ejecta. If material were being significantly
accelerated (from being hit by younger, faster material) or
decelerated (from running into older material), we would expect to see
some kind of spatial correlation of those changes in velocity, yet we
do not.  Thus, while we cannot rule out small amounts of acceleration
or deceleration for individual features, the outer ejecta appear to
be, on the whole, expanding ballistically.  We therefore treat all
velocities as constant unless otherwise noted, and use the weighted
mean as the final proper motion for each feature.

\begin{figure*}
  \centering
  \includegraphics[width=0.49\linewidth]{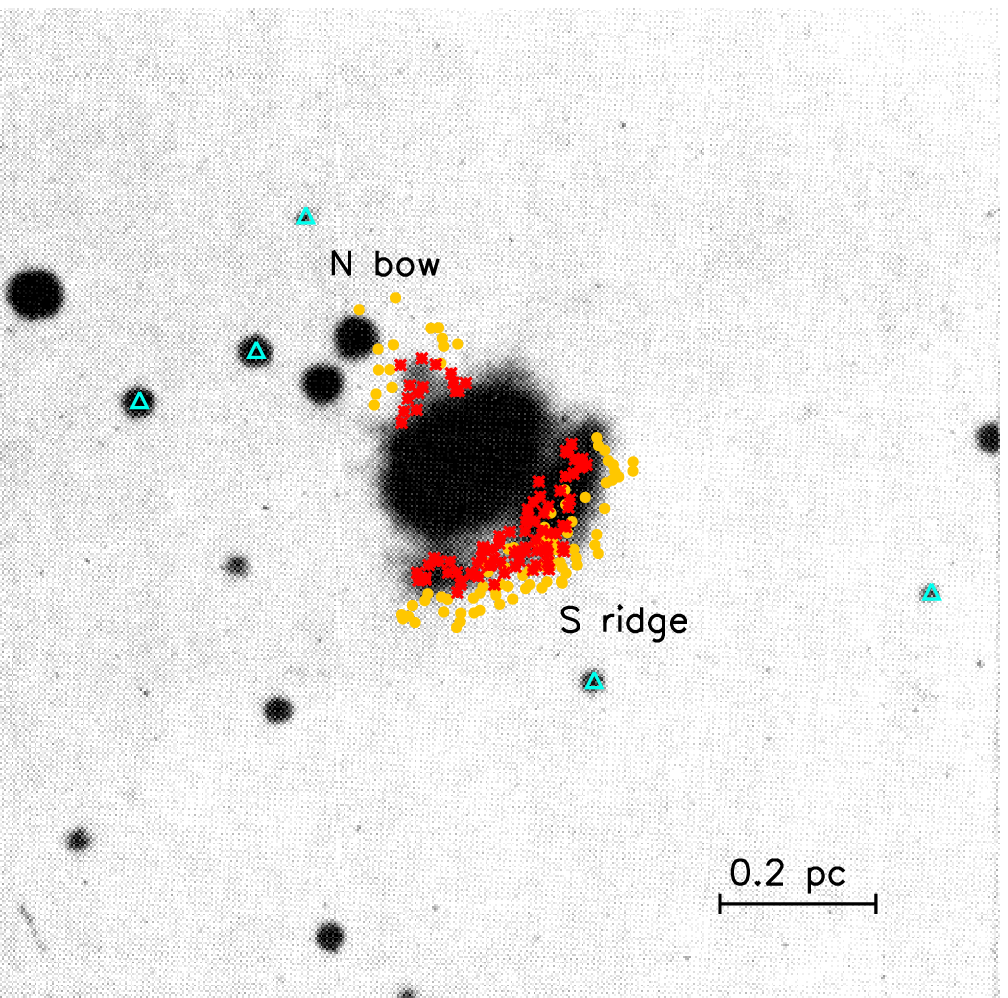}
  \includegraphics[width=0.49\linewidth]{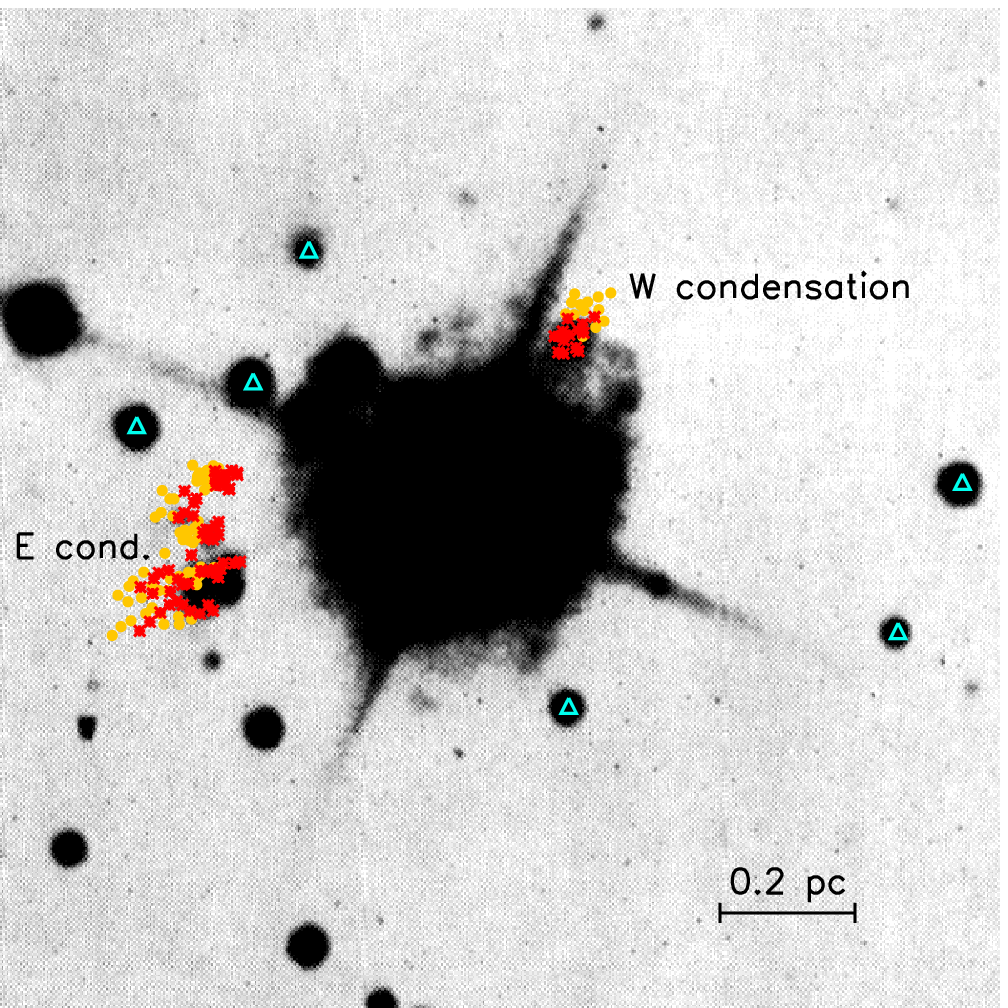}
  \caption{Left: red-sensitive photographic plate image of $\eta$ Car
    and its surrounding nebulosity from \citet{thackeray1950}.  The
    predicted 1950 positions of the features we observe in the N bow
    and S ridge are plotted as red asterisks, while our observed
    2005 positions of the same features are marked with yellow
    circles.  The stars that were used to align this image with our
    reference frame are indicated with cyan triangles.  Right: same as
    left, using a deeper image from \citet{thackeray1950} to show the
    observed current and predicted past positions of the E and W
    condensations.
  \label{fig:1950}}
\end{figure*}

As a last check on whether the assumption of constant velocities is
appropriate, we extrapolate our observed positions of the outer ejecta
back to 1950 and compare them to the ground-based photographic plate
images of \citet{thackeray1950}.  We aligned the \citet{thackeray1950}
images to a copy of our stacked ACS image that had been convolved with
a broad PSF to approximately match the resolution of the older images.
Figure \ref{fig:1950} shows the predicted 1950 positions, with the
observed 2005 positions overplotted for comparison.  There is good
agreement between our predicted positions and the observed features in
1950.  Although one must be conscious of the limitations of
photographic plate data, these data are consistent with ballistic
motion of $\eta$ Car's ejecta over the past 60 years.  \btxt{From the
  motion that we measure, however, we cannot rule out significant
  acceleration or deceleration that may have occurred earlier
  than 60 years ago.}

%----------------------------------------------------------------------
%----------------------------------------------------------------------

\subsection{Ages of the outer ejecta}
\label{subsec:ages}

\begin{figure*}
  \centering
  \includegraphics[width=0.8\linewidth, trim=0 0 0 0, clip]{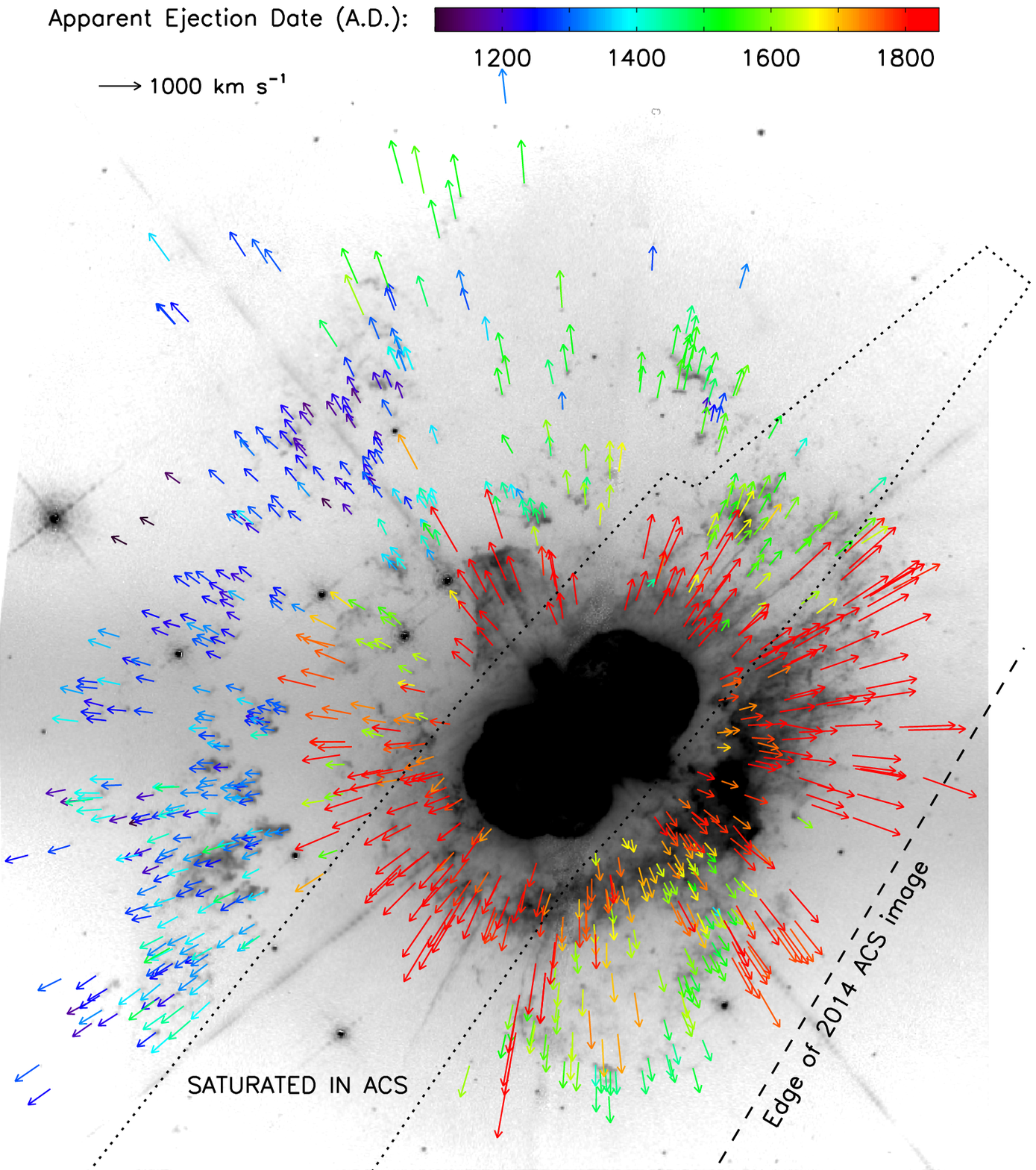}
  \caption{Vectors illustrating the observed proper motions of 792
    features in the ejecta of $\eta$ Car.  The arrows are color-coded
    by the date of ejection from the central star, calculated assuming
    constant velocity.  The region contaminated in the ACS images by
    bleeding from the highly saturated Homunculus is marked; features
    in this region were measured in WFPC2 data only.  The background
    image is the same as in Figure \ref{fig:labels}.
  \label{fig:arrows}}
\end{figure*}

\btxt{Under the assumption that the constant velocities observed over
  the last 60 years apply over the lifetime of the ejecta,} estimating
ejection dates is trivial.  \btxt{Earlier episodes of acceleration or
  deceleration would affect the specific ejection dates deduced, but,
  as we discuss in Section \ref{subsec:alt}, are unlikely to affect
  the overall relationships between groups of ejecta.  Ejection date}
results are summarized in Figure \ref{fig:arrows}: the magnitude and
direction of each knot's proper motion are indicated by the length and
orientation of its arrow, and the arrows are color-coded by apparent
ejection date.  Red arrows mark the most recently ejected material,
i.e., material that was ejected around the time of the Great Eruption,
while blue and purple arrows indicate the oldest material.  It can be
seen in Figure \ref{fig:arrows} that all 792 measured features are
moving nearly directly away from the central star.

There are several distinct groups of ejecta in Figure \ref{fig:arrows}:
\begin{enumerate}
  \item The N bow, S condensation, and S ridge (which extends nearly
    completely around the Homunculus) all date back to the early
    1800s, consistent with prior results
    \citep{walborn1978,ebbets1993,morse2001}.  As we discuss below,
    the \btxt{age of the S ridge suggests} an origin during one of the
    precursor outbursts to the Great Eruption \citep[seen as spikes in
      the historical light curve;][]{smithfrew2011}, before the
    ejection of the Homunculus.
  \item The E blobs and many of the NNE condensations date back to an
    ejection event in the thirteenth century.  This event was
    apparently highly asymmetric: there are no ejecta with this age to
    the south or west.  (Note, however, that our \emph{HST} ACS
    footprint for this reference frame cuts off in the far southwest.
    We address this issue further in Section \ref{subsec:outer}.)
  \item The SE arc, the W and NW condensations, and part of the W arc
    appear to date to an intermediate eruption in the sixteenth
    century.  In the SE and W arcs, this material is being overtaken
    by faster-moving material from the Great Eruption.  The lower S
    ridge is a mix of material of different ages, explaining why
    \citet{morse2001} dated this region to half a century before the
    Great Eruption.  

    One might surmise that the apparently intermediate ages of
    material in the SE and W arcs could result from the interaction
    between material from the 1200s and material from the 1800s.
    However, there are no newer ejecta around the NW condensation or
    the faint ejecta to the far north.  As we discuss in Section
    \ref{subsec:rv}, radial velocities also suggest that these ejecta
    are not connected to the thirteenth-century eruption.  They also
    lie outside the soft X-ray shell, which traces strong
    \btxt{current} interactions between ejecta (see Section
    \ref{subsec:xray})
\end{enumerate}

%----------------------------------------------------------------------

%----------------------------------------------------------------------

\begin{figure*}
  \centering
  \includegraphics[width=\linewidth, trim=5mm 0 5mm 0, clip]{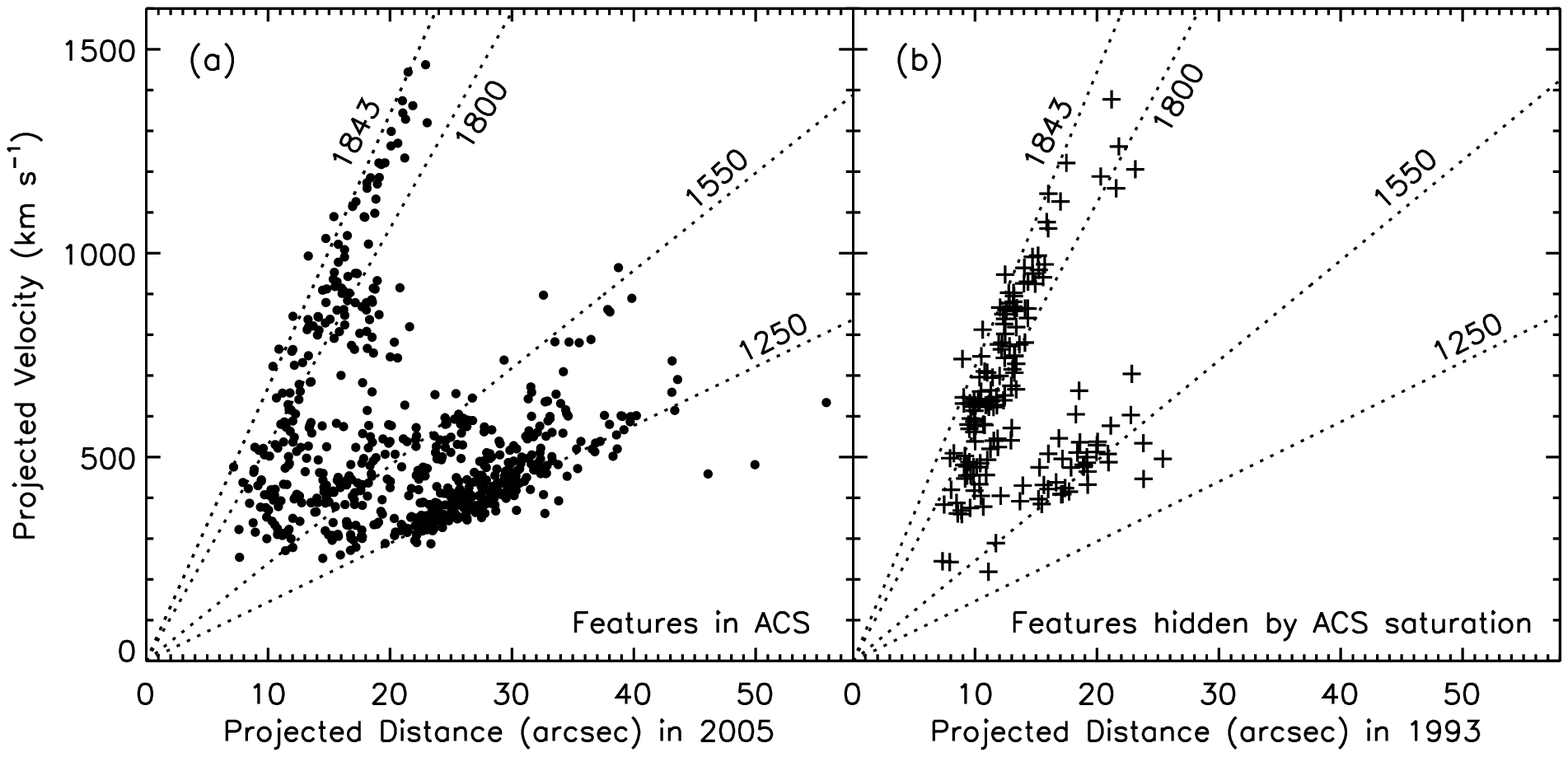}
  \caption{(a) Measured transverse velocity versus projected distance
    from the central star for all features measured over the ACS
    2005--2014 baseline.  The predicted positions of material ejected
    at the peak of the 1840s Great Eruption is marked, along with
    tracks for material ejected in 1250 A.D., 1550 A.D., and 1800
    A.D., for reference.  (b) Same as (a), but for features that were
    masked by saturation bleeds in the ACS images and were
    consequently only measured with WFPC2.
  \label{fig:vdist}}
\end{figure*}

Figure \ref{fig:vdist} presents these data in an alternative form,
plotting the transverse velocity of ejecta features versus their
projected distance from $\eta$ Car.  Again, there is evidence for at
least two eruptions, one in the early 1800s and another circa 1250
A.D.  An intermediate eruption, circa 1550 A.D., is suggested but less
obvious.  As mentioned above, the sixteenth-century ages could in
principle be the result of interaction between ejecta from the
Great Eruption and the 1200s event.  However, the intermediate-aged
features appear in a fairly clear line in velocity--distance space
\btxt{(i.e., as a Hubble-like law)} rather than being smeared between
the points from the two other events.

%----------------------------------------------------------------------
%----------------------------------------------------------------------

An additional feature of interest in Figure \ref{fig:vdist} is that
nearly all of the \btxt{nineteenth-century} features appear to have
been ejected \btxt{decades} before 1843.  Proper motions of the
Homunculus date it to the mid-1840s
\citep{currie1996,smithgehrz1998,morse2001}; this is when the light
curve of $\eta$ Car reached its peak \citep{smithfrew2011}.  There are
two possibilities for why the S ridge and its extended wings appear
older: (1) its material experienced a period of deceleration as it
interacted with circumstellar material early in its history before
continuing to expand ballistically; or (2) it was ejected prior to the
formation of the Homunculus.  \btxt{The second option is supported by
  the photometric record:} the Great Eruption was preceded by a
brightening in 1838 and a similar but poorly observed event in 1827
\citep[see][who found that these events align with the predicted
  periastron times of $\eta$ Car's binary orbit]{smithfrew2011}.  The
data before 1827 are very sparse (or nonexistent) and leave open the
possibility of additional prior periastron outbursts.  The S ridge may
have been ejected during one or more of these events rather than
during the main peak of the Great Eruption.  \btxt{Its strong
  asymmetry may therefore be related to stellar collisions at
  periastron \citep{smith2011} associated with these early
  brightenings.  The alternative explanation (an early deceleration
  episode) is plausible for the material at the edges of the bright S
  condensation, which appears in Figure \ref{fig:vdist} around
  16--20\arcsec~and 500--700 km s$^{-1}$.  We could be seeing the
  result of interaction between the S condensation (ejected during the
  Great Eruption) and the slightly older S ridge material.}

\begin{figure*}
  \centering
  \includegraphics[width=0.8\linewidth]{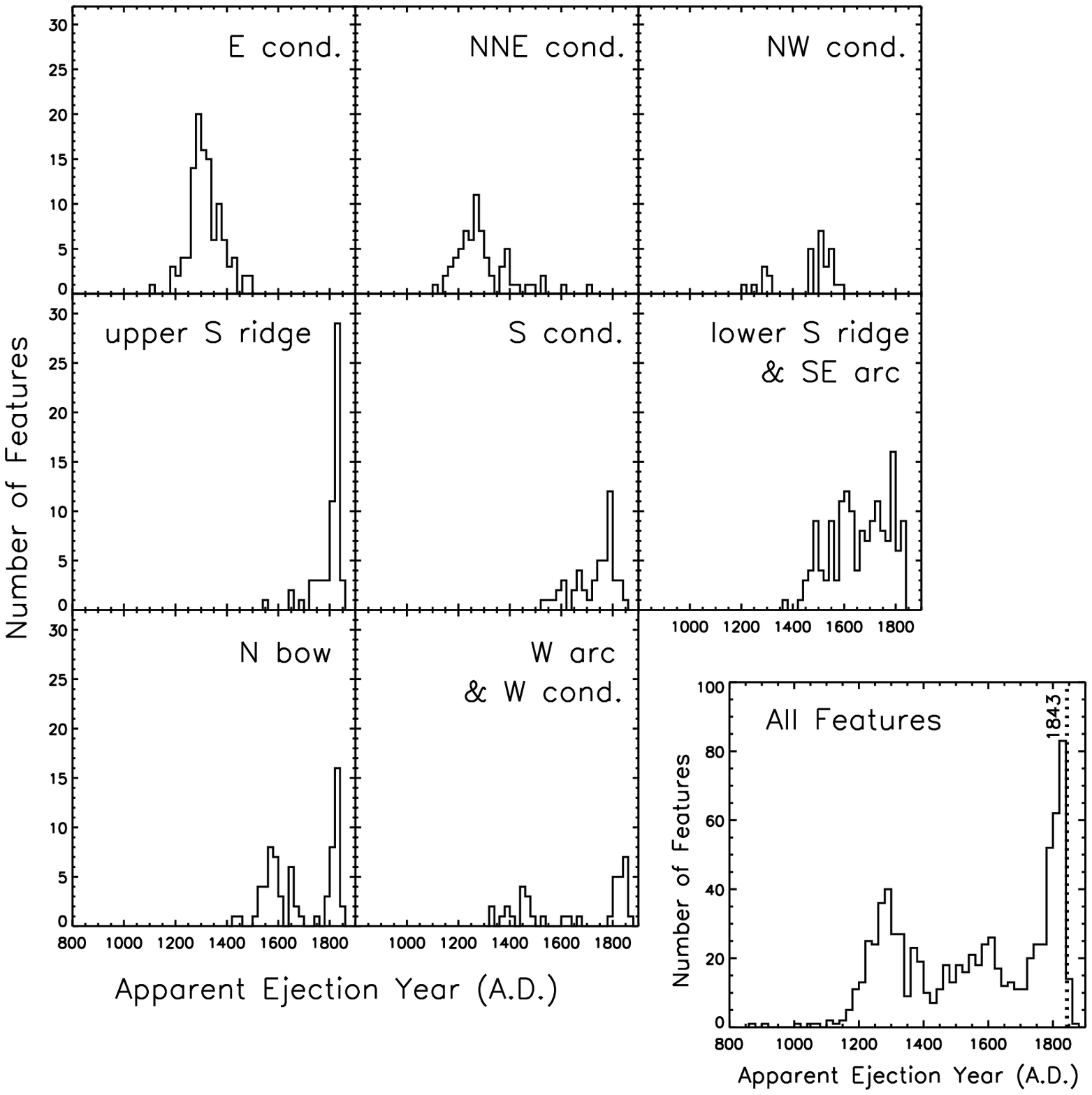}
  \caption{Histograms of the apparent ejection dates of $\eta$ Car's
    outer ejecta, assuming constant velocity since ejection.  The
    regions are as shown in Figure \ref{fig:diffimages}.  The bottom
    right panel shows the ejection date histogram for all 792 measured
    features, with the peak of the Great Eruption in 1843 marked by a
    dotted line.
  \label{fig:datehist}}
\end{figure*}

Finally, Figure \ref{fig:datehist} presents histograms of the apparent
ejection dates, both overall and for each of the regions in Figures
\ref{fig:diffimages} and \ref{fig:velhist}.  The differences in age
between the various groups of features are clearly evident.  The E and
NNE condensations, for instance, are much older than the upper S ridge
and S condensations.  The mix of ages in the W arc and the lower S
ridge is also apparent.  As in Figure \ref{fig:vdist}, the overall
histogram of ejection dates shows two obvious distinct events, in the
thirteenth and nineteenth centuries, as well as a smaller peak of
features ejected in the late 1500s.

Overall, our proper motions of $\eta$ Car's outer ejecta solidly
confirm that there was at least one major mass-loss event prior to the
Great Eruption, barring a strong deceleration event occurring sometime
before 1950 \btxt{(see Section \ref{subsec:alt}).  Assuming
    constant velocities,} there is an approximately 600-year interval
  between the two most distinct eruptions.  If the third eruption is
  included, then these major mass-loss events have occurred every
  $\sim$300 years.

%----------------------------------------------------------------------
%----------------------------------------------------------------------

\subsection{Asymmetry and the most distant ejecta}
\label{subsec:outer}

\begin{figure}
  \centering
  \includegraphics[width=\columnwidth, trim=0 0 0 5mm, clip]{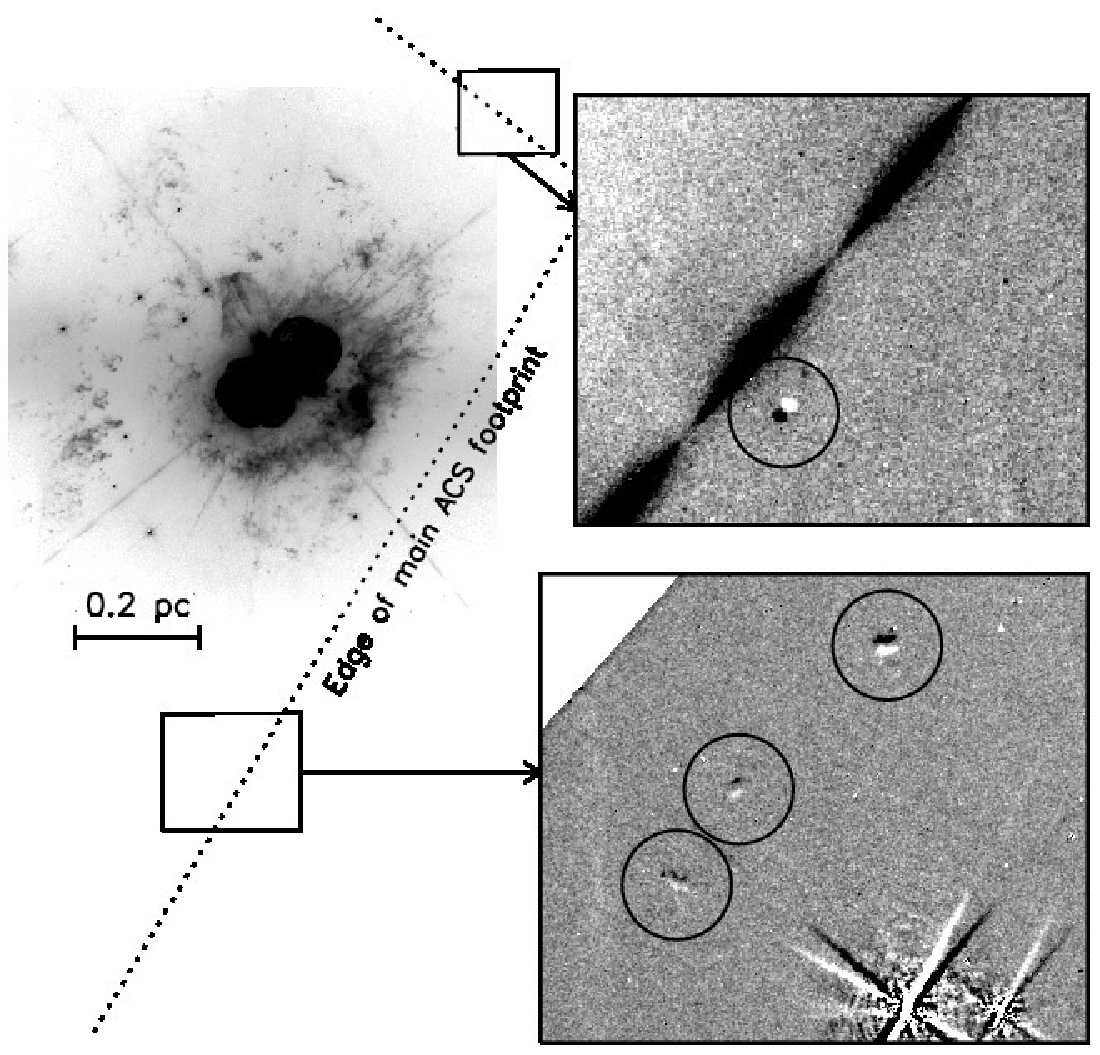}
  \caption{Difference images (ACS 2014 $-$ ACS 2005)
    showing knots identified in images outside the primary observed
    footprint.  The image from Figure \ref{fig:labels} is shown as a
    reference; the small boxes indicate where the displayed features
    are located relative to $\eta$ Car.  The four knots showing motion
    over the 9-year baseline are circled.
  \label{fig:outer}}
\end{figure}

As mentioned above and as shown in Figure \ref{fig:arrows}, the
reference frame in which we identified and measured ejecta cuts off
closer to $\eta$ Car on the southwest than on any other side,
complicating our assessment of asymmetry.  We searched the adjacent
ACS footprints \citep[which together make up a mosaic of the Tr 16
  cluster; see][]{smith2010a} for other possible ejecta.  As shown in
Figure \ref{fig:outer}, we found only four small features moving away
from $\eta$ Car, three to the south and one to the northwest.  We were
able to measure the proper motions of three of those features,
although a relative zero-point uncertainty of several km s$^{-1}$
exists between footprints because of their minimal overlap.  For the
same reason, there is also a several-pixel (100s mas) uncertainty in
distance from the central source.  

The knot to the northwest has a transverse velocity of $\sim630$ km
s$^{-1}$, giving it an estimated ejection date of 1045 $\pm$ 15 A.D.,
assuming ballistic motion.  The two knots that we were able to measure
to the far south are traveling at 460--480 km s$^{-1}$ and date to 900
$\pm$ 30 A.D.  The four outer knots could thus come from a smaller,
older mass-loss event or events. \btxt{Given} the systemic
uncertainties described above, \btxt{however,} we cannot rule out an
association with the thirteenth-century event.

Notably, there are no large-scale older ejecta similar to the E
condensations on the west side of $\eta$ Car.  It would appear that
the eruption in the 1200s was highly asymmetric, perhaps even
one-sided, sending substantial mass to the east and northeast but
little to no mass in other directions.  The Great Eruption, despite
producing the bipolar Homunculus, was also significantly asymmetric,
ejecting more mass into the extended S ridge than into the
distinctly shaped N bow.  The sixteenth-century ejecta show some
possible bipolarity, but are not aligned with the axis of the
Homunculus.

%----------------------------------------------------------------------

%----------------------------------------------------------------------

%%%%%%%%%%%%%%%%%%%%%%%%%%%%%%%%%%%%%%%%%%%%%%%%%%%%%%%%%%%%%%%%%%%%%%%
\section{Discussion}
\label{sec:disc}

\subsection{Comparison to radial velocities}
\label{subsec:rv}

To fill out the third dimension of the outer ejecta's motion, we turn
to radial velocity studies in the literature.  \citet{weis2001}
measured radial velocities of 90 distinct features in the outer ejecta,
finding speeds from 100 km s$^{-1}$ (in the SE arc) to 1960 km
s$^{-1}$ (in the upper part of the S ridge).  The E condensations, the
N bow, most of the NNE condensations, and the unnamed material on the
east side of the Homunculus---including all of the ejecta that date
back to the 1200s---are blueshifted.  In contrast, the S ridge, S
condensation, SE arc, W arc, W condensation, NW condensation, and some
of the material to the far north are redshifted.  The Homunculus
itself is also aligned in this fashion, with a blueshifted southeast
lobe and redshifted northwest lobe, tilted out of the plane of the sky
by 48$\degr$ \citep[e.g.,][]{davidson2001,smith2006b}.

The ejecta from the thirteenth-century eruption are traveling \emph{toward
us} at an angle of 20--40$\degr$ out of the plane of the sky, depending
on whether we use the radial velocities of \citet{meaburn1987} and
\citet{smithmorse2004} or those of \citet[][who present only the
  highest-magnitude velocities for each feature]{weis2001}.  The
ejecta that appear to date from the 1500s, however, are traveling \emph{away
from us}, tilted up to 30$\degr$ from the plane of the sky.  The
dividing line between the two proper motion groups on the north side
of $\eta$ Car closely follows the dividing line between blue- and
redshifted material seen by \citet{weis2001}.  With the
intermediate-aged material in a completely different part of
three-dimensional space than the oldest ejecta, it becomes much more
likely that the intermediate-aged features are from a distinct
eruption rather than the result of past acceleration/deceleration from
ejecta interactions.

\subsection{Relationship to X-ray emission \btxt{and the extremely fast ejecta}}
\label{subsec:xray}

%----------------------------------------------------------------------
\begin{figure}
  \centering
  \includegraphics[width=\columnwidth]{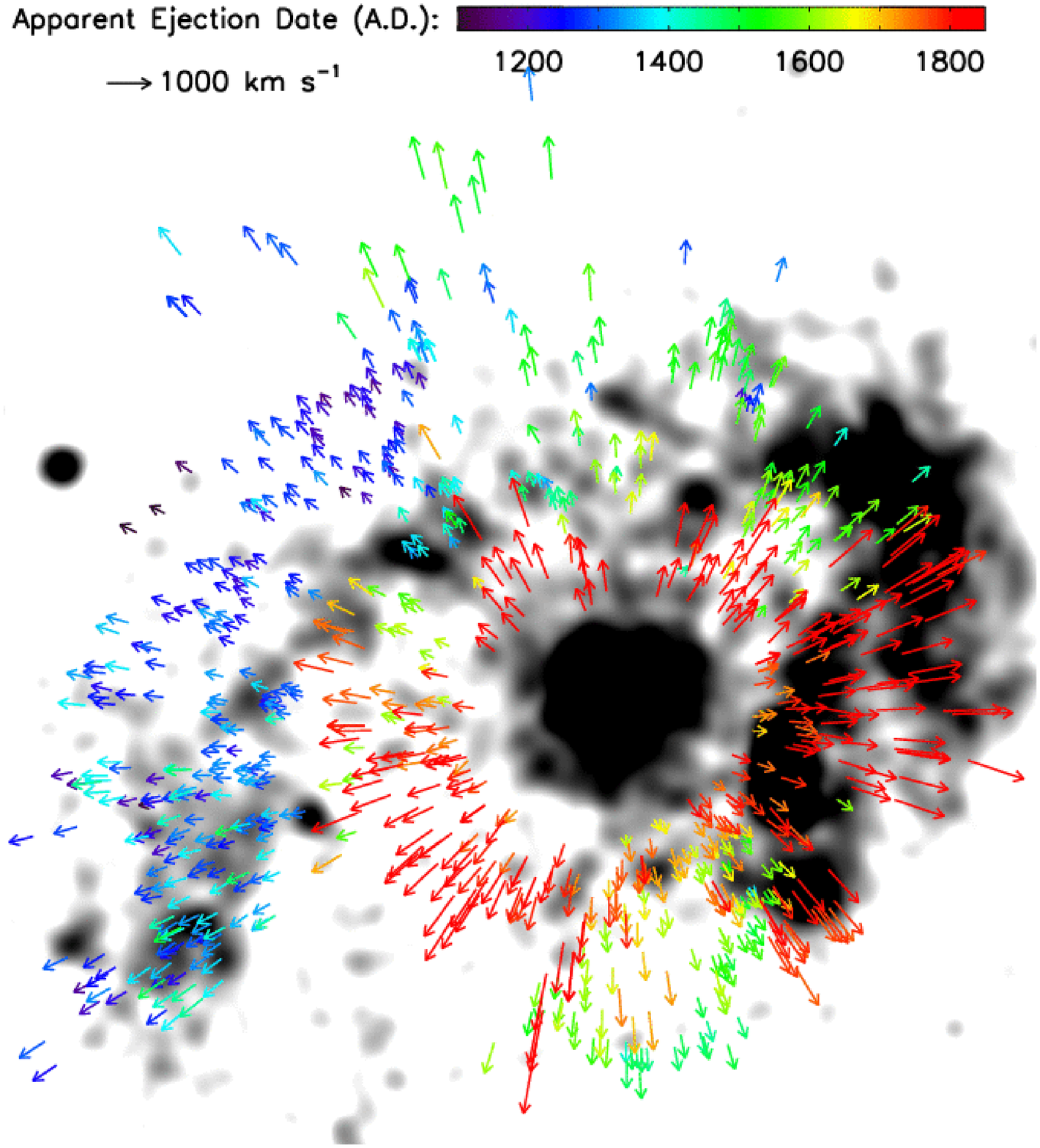}
  \caption{Similar to Figure \ref{fig:arrows}, with the scaled and
    color-coded proper motion vectors overplotted on a soft X-ray
    \emph{Chandra} image of $\eta$ Car and its surroundings.  This
    X-ray image \citep{seward2001} covers the energy range 0.5--1.5
    keV and was adaptively smoothed as in \citet{smithmorse2004}.
  \label{fig:xray}}
\end{figure}

%----------------------------------------------------------------------

$\eta$ Car is encircled by an elongated partial ring or shell of soft
X-ray emission that extends from just south of the S condensation,
over the W arc, and around to the E condensations
\citep{chlebowski1984,seward2001}. Figure \ref{fig:xray} shows a
\emph{Chandra X-ray Observatory} image of $\eta$ Car and its
surroundings \citep{seward2001}, with the proper motion velocity
vectors for the optical ejecta plotted on top.  The X-ray shell has
roughly the same axis orientation as the Homunculus, but has a notable gap
to the south and southeast.

The spatial association between the soft X-ray shell and the optical
features in the outer ejecta has led to a broad consensus that the
X-ray emission is the result of shock heating from ejecta interactions
\citep{chlebowski1984,corcoran1998,seward2001,weis2001,smithmorse2004}.
The strongest soft X-ray emission is coincident with the S
condensation, upper S ridge, and W arc, where we have measured a mix
of ages for the dense optical ejecta.  Extremely fast ejecta, with
radial speeds of up to $\sim3000$ km s$^{-1}$, have been detected
close to the E condensations and coincident with the X-ray shell
\citep{smithmorse2004,smith2008}.  \btxt{As described in Section
  \ref{subsec:wfpc2},} this very fast material, originating from the
Great Eruption in the nineteenth century, is Doppler-shifted out of
the narrowband filters used here \btxt{and is not detected in our
  images.  It} is presumably interacting with \btxt{or approaching}
the denser, slower blobs measured in this paper.  \btxt{Additional
  spectral mapping of the outer ejecta is needed to explore the full
  distribution of the high-velocity material and its relationship to
  the soft X-ray emission.}

Given the limits on the deceleration that we
measured for the knots in the E condensations, we can roughly estimate
their relative density compared to the very fast ejecta.  Over 21
years of \emph{HST} observations, the features in the E condensations
change velocity by an average of -0.1 $\pm$ 1.2 km s$^{-1}$ yr$^{-1}$
(consistent with zero).  Assuming that the collision between ejecta is
fully inelastic and that momentum is conserved, the observed E
condensations are thus 10--10$^4$ times denser than the impacting very
fast ejecta, depending on the timescale over which the fast ejecta
decelerate.

Returning to Figure \ref{fig:xray}, note that there is no
\btxt{spatial} correlation between the sixteenth-century ejecta and
the soft X-ray emission.  The SE arc falls in the X-ray gap, and the
northern ejecta lie outside the X-ray shell altogether.  If the
intermediate ages of these ejecta were the result of interaction
between the thirteenth- and nineteenth-century ejecta, we would expect
to see strong X-ray emission from the location of these interactions.
We would also expect there to be apparently sixteenth-century ejecta
around the E blobs, where very fast ejecta are observed to be hitting
the thirteenth-century material.  We observe neither of these things,
further strengthening the case that there was a distinct eruption in the
1500s.

\subsection{Alternate ejection histories}
\label{subsec:alt}

\btxt{In our analysis, we have found approximate ejection dates by
  assuming that there was no substantial, large-scale deceleration (or
  acceleration) of the outer ejecta prior to their first detection in
  1950.  $\eta$ Car is surrounded by a ``cocoon'' of
  less-nitrogen-rich material, likely from pre-eruption mass loss
  \citep{smithmorse2004}.  Here, we explore the hypothesis that all of
  the observed outer ejecta were decelerated early in their history as
  they encountered this material, then tapered to their current
  coasting velocities as the density of the cocoon diminished with
  radius.  In this case, the ejection dates of all the knots move
  later and closer together.  The true timescale between eruptions
  could then be closer to 100--200 years instead of 300 years.
  However, our main result, the detection of multiple major mass-loss
  events in $\eta$ Car's history, is unaffected.  That is the most
  straightforward scenario to explain our observations.}

\btxt{The most extreme alternate ejection history---that all of the
  observed outer ejecta date to the Great Eruption of the 1840s, but
  suffered different amounts of deceleration with an asymmetric
  wind---is much less likely.  It requires a complex pre-existing
  configuration of mass to decelerate the observed outer ejecta such
  that they reach their current coasting velocities at their current
  distances from $\eta$ Car.  A spherically symmetric distribution of
  pre-eruption circumstellar material (e.g., a stellar wind with a
  $r^{-2}$ density profile) could not have produced the different
  coasting velocities seen at the same radii from the star.  Compare,
  for example, the NNE and NW condensations: although both are
  $\sim$25\arcsec~from $\eta$ Car, the average proper motion of the
  NNE knots is 380 km s$^{-1}$ while the NW knots are moving at 550 km
  s$^{-1}$.}

\btxt{The E condensations provide a particularly firm constraint, as
  they are already 19--23\arcsec~from the central star in the
  \citet{thackeray1950} images.  To reach that distance by 1950 with
  an origin in the 1840s would require a minimum initial projected
  speed of $\sim$2000 km s$^{-1}$.  The measured projected speeds of
  the E condensations over the past two decades are 350--600 km
  s$^{-1}$.  If these features were ejected in the 1840s, there is no
  way for them to have reached the observed separation from $\eta$ Car
  and yet have their current, relatively low speeds without invoking a
  dense shell located close to their 1950 positions.  One could then
  posit additional pre-existing partial shells that would have
  decelerated the NW and SE condensations to their present-day
  coasting velocities.  This scenario would still require substantial
  and episodic prior mass-loss events.  It would have much the same
  consequences for $\eta$ Car's eruptive history, except it would
  indicate that the star's surface nitrogen enhancement took place
  recently, between the older shell-producing events and the 1840s
  Great Eruption.}

\btxt{From an energy standpoint, we know the Great Eruption did
  produce ejecta with velocities of 1000 km s$^{-1}$ or more; they are
  seen in spectra of the equatorial plane of the Homunculus and of
  near or inside the soft X-ray shell \citep{smith2008}.  However, as
  discussed in Section \ref{subsec:xray}, some of these very fast
  ejecta are coincident with and appear to be running into the E
  condensations, likely powering the observed soft X-ray shell.  There
  is no clear mechanism by which two sets of mass should reach the
  same projected distance over the same time yet have present-day
  speeds that differ by a factor of ten.  Furthermore, the ejection of
  all of the dense outer ejecta at 1000 km s$^{-1}$ or more would
  require a dramatic upward revision of the estimated energy budget of
  the Great Eruption.  The discovery of the very fast ejecta roughly
  doubled the Great Eruption's energy budget \citep{smith2008}, and
  that number assumed that the 3000 km s$^{-1}$ material was about a
  factor of ten less dense than the slower, named ejecta.  (Note that
  this agrees with our rough estimate of the density ratio in Section
  \ref{subsec:xray}.)  The explosive amount of kinetic energy required
  to eject several solar masses of outer ejecta at these speeds
  approaches that of a core-collapse supernova.}

\subsection{Implications for models of Eta Car}
\label{subsec:hist}

The proper motions of the outer ejecta \btxt{thus} confirm that $\eta$
Car has undergone at least two and probably three major mass-loss
events over the last millennium, including the nineteenth-century Great
Eruption.  These results raise a number of interesting questions that
are as yet unanswered by existing theories: What mechanism drives
these repeating eruptions?  Why do they repeat on a
\btxt{several-hundred}-year timescale, much longer than the 5.5-year
binary orbit?  Why was the thirteenth-century eruption so one-sided?
Why does the Homunculus have a clear bipolar symmetry that past
eruptions lack?  Models of $\eta$ Car's behavior must not treat the
Great Eruption in isolation, but should aim to account for all of the
observed characteristics of the prior eruptions.

In single-star models (i.e., those in which the companion star plays
no significant role), $\eta$ Car's eruptions are considered to be
continuum-driven super-Eddington wind events, where the extra
luminosity arises from as-yet-unidentified mechanisms
\citep{davidsonhumphreys1997,owocki2004}.  However, it is difficult
for a single-star model to explain the observed asymmetry in the outer
ejecta and the changes in mass-loss symmetry with time.  What could
cause a single star to produce the highly one-sided mass loss of the
thirteenth-century eruption, then eject the axisymmetric Homunculus
several centuries later?

Current models involving a binary or higher-order multiple system
\btxt{may get us closer, but} also have difficulty explaining the
observed historical mass loss.  The time between major eruptive events
is much longer than the orbital period of the current binary
\citep[5.54 years,][]{damineli1996}.  If the eruptions are influenced
by periastron interactions
\citep{cassinelli1999,soker2007,kashisoker2010,smith2011}, these
interactions must be suppressed \btxt{for long periods of} time.
Models that invoke a one-time catastrophic event such as a merger
\citep{gallagher1989,iben1999,morrispodsiadlowski2009,podsiadlowski2010,portegieszwartvandenheuvel2016}
or a dynamical exchange \citep{liviopringle1998} require additional
unexplained complexities to account for these repetitive yet discrete
events separated by centuries.

The lack of older, more distant outer ejecta rules out additional
major mass-loss events prior to the thirteenth century.
\citet{bohigas2000} claimed to detect a much older (10$^{4}$ yr)
bipolar shell, but this was interpreted by \citet{smith2005b} as
$\eta$ Car's astropause, \btxt{modified by the wind of a nearby Of
  star,} as it lacks the nitrogen-rich chemistry or clumpy structure
of the other outer ejecta.  What, then, caused $\eta$ Car's eruptive
behavior to start in the 1200s?  The apparently sudden initiation of
mass ejections is probably an important clue, and is another key
constraint for models.

An added complication in the story of $\eta$ Car is the 1890s Lesser
Eruption, in which a much smaller amount of mass \btxt{\citep[0.1
    M$_{\sun}$;][]{ishibashi2003,smith2005a}} was ejected with the
same geometry as the Homunculus.  Secondary eruptions of this sort,
occurring several decades after a major outburst, have been observed in
P Cygni and some of the other known eruptive LBVs
\citep{humphreys1999}.  Our data are not sensitive to similar small
eruptions in the decades after the larger thirteenth- and
sixteenth-century events.

%%%%%%%%%%%%%%%%%%%%%%%%%%%%%%%%%%%%%%%%%%%%%%%%%%%%%%%%%%%%%%%%%%%%%%%
\section{Conclusions}
\label{sec:conc}

We have aligned eight epochs of \emph{HST} imaging (both WFPC2 and
ACS) of $\eta$ Car's outer ejecta to the same distortion-corrected
reference frame and measured the proper motions of 792 ejecta
features, many for the first time.  We achieve unprecedented time
coverage, with each feature measured in up to 16 baselines over 21
years, as well as unprecedented velocity precision (few km s$^{-1}$)
and spatial resolution for these features.

All 792 features measured in $\eta$ Car's outer ejecta have transverse
velocities pointing nearly directly away from the star.  The majority
\btxt{have proper motions} of 300--600 km s$^{-1}$, although some are
as fast as 1500 km s$^{-1}$.  The fastest-moving material is found in
the large feature known as the S ridge and in the broadly
jet-shaped N bow.  Both date back to $\eta$ Car's Great Eruption in
the 1840s or to a few decades prior.

Over the 21 years of data, we see no evidence for large-scale
acceleration or deceleration of any of the outer ejecta: 94\% of the
knots are consistent with moving at constant velocity over that time.
\btxt{Comparison to images from 1949--1950 support ballistic motion
  over a longer time period.}  Under the assumption of constant
velocity, we find that the material in and around the E and NNE
condensations was ejected in the mid-1200s A.D., give or take 50--100
years.  With the exception of three small knots to the far south of
$\eta$ Car and one to the northwest, the ejecta dating to the
mid-1200s are all found to one side of the central star and are
blueshifted.

We also see evidence of a third, intermediate eruption that occurred
in the sixteenth century.  Ejecta dating to the mid-1500s are found in
the SE arc, the W condensation, and in and around the NW condensation.
From proper motions alone, we cannot rule out that this intermediate
date peak is the result of newer ejecta from the Great Eruption
hitting the older material from the 1200s.  However, the radial
velocities of these features place them in a different part of
three-dimensional space from the thirteenth-century ejecta.  The lack
of X-ray emission over the SE arc and the features to the far north
also indicates a lack of strong interaction between ejecta at those
spots.

In summary, we have shown with distance-independent measurements that
$\eta$ Car erupted at least once, likely twice, before its Great
Eruption in the 1800s.  Models for this still-enigmatic star must
therefore explain the recurrence of these major mass-loss events,
along with their \btxt{several-hundred-}year timescale and their
various asymmetries.

%%%%%%%%%%%%%%%%%%%%%%%%%%%%%%%%%%%%%%%%%%%%%%%%%%%%%%%%%%%%%%%%%%%%%%%
\section*{Acknowledgements}

The authors would like to thank Jay Anderson for providing us with his
suite of PSF-fitting and image alignment software, and for his
valuable instruction, guidance, and technical support.  \btxt{We also
  thank the anonymous referee for constructive comments that improved
  the paper.}  Support for this work was provided by NASA grant
GO-13390 from the Space Telescope Science Institute, which is operated
by the Association of Universities for Research in Astronomy,
Inc. under NASA contract NAS 5-26555.  This work is based on
observations made with the NASA/ESA \emph{Hubble Space Telescope},
obtained from the Data Archive at the Space Telescope Science
Institute.

%%%%%%%%%%%%%%%%%%%%%%%%%%%%%%%%%%%%%%%%%%%%%%%%%%%%%%%%%%%%%%%%%%%%%%%
\bibliographystyle{mnras}
\bibliography{ms} 

\begin{thebibliography}{}
\makeatletter
\relax
\def\mn@urlcharsother{\let\do\@makeother \do\$\do\&\do\#\do\^\do\_\do\%\do\~}
\def\mn@doi{\begingroup\mn@urlcharsother \@ifnextchar [ {\mn@doi@}
  {\mn@doi@[]}}
\def\mn@doi@[#1]#2{\def\@tempa{#1}\ifx\@tempa\@empty \href
  {http://dx.doi.org/#2} {doi:#2}\else \href {http://dx.doi.org/#2} {#1}\fi
  \endgroup}
\def\mn@eprint#1#2{\mn@eprint@#1:#2::\@nil}
\def\mn@eprint@arXiv#1{\href {http://arxiv.org/abs/#1} {{\tt arXiv:#1}}}
\def\mn@eprint@dblp#1{\href {http://dblp.uni-trier.de/rec/bibtex/#1.xml}
  {dblp:#1}}
\def\mn@eprint@#1:#2:#3:#4\@nil{\def\@tempa {#1}\def\@tempb {#2}\def\@tempc
  {#3}\ifx \@tempc \@empty \let \@tempc \@tempb \let \@tempb \@tempa \fi \ifx
  \@tempb \@empty \def\@tempb {arXiv}\fi \@ifundefined
  {mn@eprint@\@tempb}{\@tempb:\@tempc}{\expandafter \expandafter \csname
  mn@eprint@\@tempb\endcsname \expandafter{\@tempc}}}

\bibitem[\protect\citeauthoryear{{Anderson}}{{Anderson}}{2006}]{anderson2006}
{Anderson} J.,  2006, in {Koekemoer} A.~M.,  {Goudfrooij} P.,   {Dressel}
  L.~L.,  eds, The 2005 HST Calibration Workshop: Hubble After the Transition
  to Two-Gyro Mode. p.~11

\bibitem[\protect\citeauthoryear{{Anderson} \& {King}}{{Anderson} \&
  {King}}{1999}]{andersonking1999}
{Anderson} J.,  {King} I.~R.,  1999, \mn@doi [\pasp] {10.1086/316432}, \href
  {http://adsabs.harvard.edu/abs/1999PASP..111.1095A} {111, 1095}

\bibitem[\protect\citeauthoryear{{Anderson} \& {King}}{{Anderson} \&
  {King}}{2000}]{andersonking2000}
{Anderson} J.,  {King} I.~R.,  2000, \mn@doi [\pasp] {10.1086/316632}, \href
  {http://adsabs.harvard.edu/abs/2000PASP..112.1360A} {112, 1360}

\bibitem[\protect\citeauthoryear{{Anderson} \& {King}}{{Anderson} \&
  {King}}{2006}]{andersonking2006}
{Anderson} J.,  {King} I.~R.,  2006, Technical report, {PSFs, Photometry, and
  Astronomy for the ACS/WFC}.
Space Telescope Science Institute

\bibitem[\protect\citeauthoryear{{Anderson} \& {van der Marel}}{{Anderson} \&
  {van der Marel}}{2010}]{andersonvandermarel2010}
{Anderson} J.,  {van der Marel} R.~P.,  2010, \mn@doi [\apj]
  {10.1088/0004-637X/710/2/1032}, \href
  {http://adsabs.harvard.edu/abs/2010ApJ...710.1032A} {710, 1032}

\bibitem[\protect\citeauthoryear{{Anderson} et~al.,}{{Anderson}
  et~al.}{2008a}]{anderson2008a}
{Anderson} J.,  et~al., 2008a, \mn@doi [\aj] {10.1088/0004-6256/135/6/2055},
  \href {http://adsabs.harvard.edu/abs/2008AJ....135.2055A} {135, 2055}

\bibitem[\protect\citeauthoryear{{Anderson} et~al.,}{{Anderson}
  et~al.}{2008b}]{anderson2008b}
{Anderson} J.,  et~al., 2008b, \mn@doi [\aj] {10.1088/0004-6256/135/6/2114},
  \href {http://adsabs.harvard.edu/abs/2008AJ....135.2114A} {135, 2114}

\bibitem[\protect\citeauthoryear{{Bohigas}, {Tapia}, {Ruiz}  \&
  {Roth}}{{Bohigas} et~al.}{2000}]{bohigas2000}
{Bohigas} J.,  {Tapia} M.,  {Ruiz} M.~T.,   {Roth} M.,  2000, \mn@doi [\mnras]
  {10.1046/j.1365-8711.2000.03217.x}, \href
  {http://adsabs.harvard.edu/abs/2000MNRAS.312..295B} {312, 295}

\bibitem[\protect\citeauthoryear{{Cassinelli}}{{Cassinelli}}{1999}]{cassinelli1999}
{Cassinelli} J.~P.,  1999, in {Morse} J.~A.,  {Humphreys} R.~M.,   {Damineli}
  A.,  eds,  Astronomical Society of the Pacific Conference Series Vol. 179,
  Eta Carinae at The Millennium. p.~358

\bibitem[\protect\citeauthoryear{{Chlebowski}, {Seward}, {Swank}  \&
  {Szymkowiak}}{{Chlebowski} et~al.}{1984}]{chlebowski1984}
{Chlebowski} T.,  {Seward} F.~D.,  {Swank} J.,   {Szymkowiak} A.,  1984,
  \mn@doi [\apj] {10.1086/162143}, \href
  {http://adsabs.harvard.edu/abs/1984ApJ...281..665C} {281, 665}

\bibitem[\protect\citeauthoryear{{Corcoran} et~al.,}{{Corcoran}
  et~al.}{1998}]{corcoran1998}
{Corcoran} M.~F.,  et~al., 1998, \mn@doi [\apj] {10.1086/305190}, \href
  {http://adsabs.harvard.edu/abs/1998ApJ...494..381C} {494, 381}

\bibitem[\protect\citeauthoryear{{Corcoran}, {Ishibashi}, {Swank}  \&
  {Petre}}{{Corcoran} et~al.}{2001}]{corcoran2001}
{Corcoran} M.~F.,  {Ishibashi} K.,  {Swank} J.~H.,   {Petre} R.,  2001, \mn@doi
  [\apj] {10.1086/318416}, \href
  {http://adsabs.harvard.edu/abs/2001ApJ...547.1034C} {547, 1034}

\bibitem[\protect\citeauthoryear{{Currie} et~al.,}{{Currie}
  et~al.}{1996}]{currie1996}
{Currie} D.~G.,  et~al., 1996, \mn@doi [\aj] {10.1086/118083}, \href
  {http://adsabs.harvard.edu/abs/1996AJ....112.1115C} {112, 1115}

\bibitem[\protect\citeauthoryear{{Currie}, {Dorland}  \& {Kaufer}}{{Currie}
  et~al.}{2002}]{currie2002}
{Currie} D.~G.,  {Dorland} B.~N.,   {Kaufer} A.,  2002, \mn@doi [\aap]
  {10.1051/0004-6361:20020805}, \href
  {http://adsabs.harvard.edu/abs/2002A%26A...389L..65C} {389, L65}

\bibitem[\protect\citeauthoryear{{Damineli}}{{Damineli}}{1996}]{damineli1996}
{Damineli} A.,  1996, \mn@doi [\apjl] {10.1086/309961}, \href
  {http://adsabs.harvard.edu/abs/1996ApJ...460L..49D} {460, L49}

\bibitem[\protect\citeauthoryear{{Damineli}, {Conti}  \& {Lopes}}{{Damineli}
  et~al.}{1997}]{damineli1997}
{Damineli} A.,  {Conti} P.~S.,   {Lopes} D.~F.,  1997, \mn@doi [\na]
  {10.1016/S1384-1076(97)00008-0}, \href
  {http://adsabs.harvard.edu/abs/1997NewA....2..107D} {2, 107}

\bibitem[\protect\citeauthoryear{{Damineli}, {Kaufer}, {Wolf}, {Stahl}, {Lopes}
   \& {de Ara{\'u}jo}}{{Damineli} et~al.}{2000}]{damineli2000}
{Damineli} A.,  {Kaufer} A.,  {Wolf} B.,  {Stahl} O.,  {Lopes} D.~F.,   {de
  Ara{\'u}jo} F.~X.,  2000, \mn@doi [\apjl] {10.1086/312441}, \href
  {http://adsabs.harvard.edu/abs/2000ApJ...528L.101D} {528, L101}

\bibitem[\protect\citeauthoryear{{Davidson}}{{Davidson}}{1971}]{davidson1971}
{Davidson} K.,  1971, \mn@doi [\mnras] {10.1093/mnras/154.4.415}, \href
  {http://adsabs.harvard.edu/abs/1971MNRAS.154..415D} {154, 415}

\bibitem[\protect\citeauthoryear{{Davidson} \& {Humphreys}}{{Davidson} \&
  {Humphreys}}{1997}]{davidsonhumphreys1997}
{Davidson} K.,  {Humphreys} R.~M.,  1997, \mn@doi [\araa]
  {10.1146/annurev.astro.35.1.1}, \href
  {http://adsabs.harvard.edu/abs/1997ARA%26A..35....1D} {35, 1}

\bibitem[\protect\citeauthoryear{{Davidson}, {Dufour}, {Walborn}  \&
  {Gull}}{{Davidson} et~al.}{1986}]{davidson1986}
{Davidson} K.,  {Dufour} R.~J.,  {Walborn} N.~R.,   {Gull} T.~R.,  1986,
  \mn@doi [\apj] {10.1086/164301}, \href
  {http://adsabs.harvard.edu/abs/1986ApJ...305..867D} {305, 867}

\bibitem[\protect\citeauthoryear{{Davidson}, {Smith}, {Gull}, {Ishibashi}  \&
  {Hillier}}{{Davidson} et~al.}{2001}]{davidson2001}
{Davidson} K.,  {Smith} N.,  {Gull} T.~R.,  {Ishibashi} K.,   {Hillier} D.~J.,
  2001, \mn@doi [\aj] {10.1086/319419}, \href
  {http://adsabs.harvard.edu/abs/2001AJ....121.1569D} {121, 1569}

\bibitem[\protect\citeauthoryear{{Ebbets}, {Malumuth}, {Davidson}, {White}  \&
  {Walborn}}{{Ebbets} et~al.}{1993}]{ebbets1993}
{Ebbets} D.,  {Malumuth} E.,  {Davidson} K.,  {White} R.,   {Walborn} N.,
  1993, in {Cassinelli} J.~P.,  {Churchwell} E.~B.,  eds,  Astronomical Society
  of the Pacific Conference Series Vol. 35, Massive Stars: Their Lives in the
  Interstellar Medium. p.~263

\bibitem[\protect\citeauthoryear{{Frew}}{{Frew}}{2004}]{frew2004}
{Frew} D.~J.,  2004, Journal of Astronomical Data, \href
  {http://adsabs.harvard.edu/abs/2004JAD....10....6F} {10, 6}

\bibitem[\protect\citeauthoryear{{Gallagher}}{{Gallagher}}{1989}]{gallagher1989}
{Gallagher} J.~S.,  1989, in {Davidson} K.,  {Moffat} A.~F.~J.,   {Lamers}
  H.~J.~G.~L.~M.,  eds,  Astrophysics and Space Science Library Vol. 157, IAU
  Colloq. 113: Physics of Luminous Blue Variables. pp 185--192

\bibitem[\protect\citeauthoryear{{Glatzel}}{{Glatzel}}{1994}]{glatzel1994}
{Glatzel} W.,  1994, \mn@doi [\mnras] {10.1093/mnras/271.1.66}, \href
  {http://adsabs.harvard.edu/abs/1994MNRAS.271...66G} {271}

\bibitem[\protect\citeauthoryear{{Glatzel} \& {Kiriakidis}}{{Glatzel} \&
  {Kiriakidis}}{1993}]{glatzelkiriakidis1993}
{Glatzel} W.,  {Kiriakidis} M.,  1993, \mn@doi [\mnras]
  {10.1093/mnras/263.2.375}, \href
  {http://adsabs.harvard.edu/abs/1993MNRAS.263..375G} {263, 375}

\bibitem[\protect\citeauthoryear{{Gomez}, {Vlahakis}, {Stretch}, {Dunne},
  {Eales}, {Beelen}, {Gomez}  \& {Edmunds}}{{Gomez} et~al.}{2010}]{gomez2010}
{Gomez} H.~L.,  {Vlahakis} C.,  {Stretch} C.~M.,  {Dunne} L.,  {Eales} S.~A.,
  {Beelen} A.,  {Gomez} E.~L.,   {Edmunds} M.~G.,  2010, \mn@doi [\mnras]
  {10.1111/j.1745-3933.2009.00784.x}, \href
  {http://adsabs.harvard.edu/abs/2010MNRAS.401L..48G} {401, L48}

\bibitem[\protect\citeauthoryear{{Hartigan}, {Morse}, {Reipurth}, {Heathcote}
  \& {Bally}}{{Hartigan} et~al.}{2001}]{hartigan2001}
{Hartigan} P.,  {Morse} J.~A.,  {Reipurth} B.,  {Heathcote} S.,   {Bally} J.,
  2001, \mn@doi [\apjl] {10.1086/323976}, \href
  {http://adsabs.harvard.edu/abs/2001ApJ...559L.157H} {559, L157}

\bibitem[\protect\citeauthoryear{{Humphreys} \& {Davidson}}{{Humphreys} \&
  {Davidson}}{1994}]{humphreysdavidson1994}
{Humphreys} R.~M.,  {Davidson} K.,  1994, \mn@doi [\pasp] {10.1086/133478},
  \href {http://adsabs.harvard.edu/abs/1994PASP..106.1025H} {106, 1025}

\bibitem[\protect\citeauthoryear{{Humphreys}, {Davidson}  \&
  {Smith}}{{Humphreys} et~al.}{1999}]{humphreys1999}
{Humphreys} R.~M.,  {Davidson} K.,   {Smith} N.,  1999, \mn@doi [\pasp]
  {10.1086/316420}, \href {http://adsabs.harvard.edu/abs/1999PASP..111.1124H}
  {111, 1124}

\bibitem[\protect\citeauthoryear{{Iben}}{{Iben}}{1999}]{iben1999}
{Iben} Jr. I.,  1999, in {Morse} J.~A.,  {Humphreys} R.~M.,   {Damineli} A.,
  eds,  Astronomical Society of the Pacific Conference Series Vol. 179, Eta
  Carinae at The Millennium. p.~367

\bibitem[\protect\citeauthoryear{{Innes}}{{Innes}}{1903}]{innes1903}
{Innes} R.~T.~A.,  1903, Annals of the Cape Observatory, 9, 75B

\bibitem[\protect\citeauthoryear{{Ishibashi} et~al.,}{{Ishibashi}
  et~al.}{2003}]{ishibashi2003}
{Ishibashi} K.,  et~al., 2003, \mn@doi [\aj] {10.1086/375306}, \href
  {http://adsabs.harvard.edu/abs/2003AJ....125.3222I} {125, 3222}

\bibitem[\protect\citeauthoryear{{Kashi} \& {Soker}}{{Kashi} \&
  {Soker}}{2010}]{kashisoker2010}
{Kashi} A.,  {Soker} N.,  2010, \mn@doi [\apj] {10.1088/0004-637X/723/1/602},
  \href {http://adsabs.harvard.edu/abs/2010ApJ...723..602K} {723, 602}

\bibitem[\protect\citeauthoryear{{Lamers} \& {Fitzpatrick}}{{Lamers} \&
  {Fitzpatrick}}{1988}]{lamersfitzpatrick1988}
{Lamers} H.~J.~G.~L.~M.,  {Fitzpatrick} E.~L.,  1988, \mn@doi [\apj]
  {10.1086/165894}, \href {http://adsabs.harvard.edu/abs/1988ApJ...324..279L}
  {324, 279}

\bibitem[\protect\citeauthoryear{{Livio} \& {Pringle}}{{Livio} \&
  {Pringle}}{1998}]{liviopringle1998}
{Livio} M.,  {Pringle} J.~E.,  1998, \mn@doi [\mnras]
  {10.1046/j.1365-8711.1998.01567.x}, \href
  {http://adsabs.harvard.edu/abs/1998MNRAS.295L..59L} {295, L59}

\bibitem[\protect\citeauthoryear{{Maeder}}{{Maeder}}{1983}]{maeder1983}
{Maeder} A.,  1983, \aap, \href
  {http://adsabs.harvard.edu/abs/1983A%26A...120..113M} {120, 113}

\bibitem[\protect\citeauthoryear{{Meaburn}, {Wolstencroft}  \&
  {Walsh}}{{Meaburn} et~al.}{1987}]{meaburn1987}
{Meaburn} J.,  {Wolstencroft} R.~D.,   {Walsh} J.~R.,  1987, \aap, \href
  {http://adsabs.harvard.edu/abs/1987A%26A...181..333M} {181, 333}

\bibitem[\protect\citeauthoryear{{Meaburn}, {Boumis}, {Walsh}, {Steffen},
  {Holloway}, {Williams}  \& {Bryce}}{{Meaburn} et~al.}{1996}]{meaburn1996a}
{Meaburn} J.,  {Boumis} P.,  {Walsh} J.~R.,  {Steffen} W.,  {Holloway} A.~J.,
  {Williams} R.~J.~R.,   {Bryce} M.,  1996, \mn@doi [\mnras]
  {10.1093/mnras/282.4.1313}, \href
  {http://adsabs.harvard.edu/abs/1996MNRAS.282.1313M} {282, 1313}

\bibitem[\protect\citeauthoryear{{Morris} \& {Podsiadlowski}}{{Morris} \&
  {Podsiadlowski}}{2009}]{morrispodsiadlowski2009}
{Morris} T.,  {Podsiadlowski} P.,  2009, \mn@doi [\mnras]
  {10.1111/j.1365-2966.2009.15114.x}, \href
  {http://adsabs.harvard.edu/abs/2009MNRAS.399..515M} {399, 515}

\bibitem[\protect\citeauthoryear{{Morse}}{{Morse}}{1999}]{morse1999}
{Morse} J.~A.,  1999, in {Morse} J.~A.,  {Humphreys} R.~M.,   {Damineli} A.,
  eds,  Astronomical Society of the Pacific Conference Series Vol. 179, Eta
  Carinae at The Millennium. p.~13

\bibitem[\protect\citeauthoryear{{Morse}, {Davidson}, {Bally}, {Ebbets},
  {Balick}  \& {Frank}}{{Morse} et~al.}{1998}]{morse1998}
{Morse} J.~A.,  {Davidson} K.,  {Bally} J.,  {Ebbets} D.,  {Balick} B.,
  {Frank} A.,  1998, \mn@doi [\aj] {10.1086/300581}, \href
  {http://adsabs.harvard.edu/abs/1998AJ....116.2443M} {116, 2443}

\bibitem[\protect\citeauthoryear{{Morse}, {Kellogg}, {Bally}, {Davidson},
  {Balick}  \& {Ebbets}}{{Morse} et~al.}{2001}]{morse2001}
{Morse} J.~A.,  {Kellogg} J.~R.,  {Bally} J.,  {Davidson} K.,  {Balick} B.,
  {Ebbets} D.,  2001, \mn@doi [\apjl] {10.1086/319092}, \href
  {http://adsabs.harvard.edu/abs/2001ApJ...548L.207M} {548, L207}

\bibitem[\protect\citeauthoryear{{Owocki}, {Gayley}  \& {Shaviv}}{{Owocki}
  et~al.}{2004}]{owocki2004}
{Owocki} S.~P.,  {Gayley} K.~G.,   {Shaviv} N.~J.,  2004, \mn@doi [\apj]
  {10.1086/424910}, \href {http://adsabs.harvard.edu/abs/2004ApJ...616..525O}
  {616, 525}

\bibitem[\protect\citeauthoryear{{Podsiadlowski}}{{Podsiadlowski}}{2010}]{podsiadlowski2010}
{Podsiadlowski} P.,  2010, \mn@doi [\nar] {10.1016/j.newar.2010.09.023}, \href
  {http://ukads.nottingham.ac.uk/abs/2010NewAR..54...39P} {54, 39}

\bibitem[\protect\citeauthoryear{{Portegies Zwart} \& {van den
  Heuvel}}{{Portegies Zwart} \& {van den
  Heuvel}}{2016}]{portegieszwartvandenheuvel2016}
{Portegies Zwart} S.~F.,  {van den Heuvel} E.~P.~J.,  2016, \mn@doi [\mnras]
  {10.1093/mnras/stv2787}, \href
  {http://adsabs.harvard.edu/abs/2016MNRAS.456.3401P} {456, 3401}

\bibitem[\protect\citeauthoryear{{Reiter} \& {Smith}}{{Reiter} \&
  {Smith}}{2014}]{reitersmith2014}
{Reiter} M.,  {Smith} N.,  2014, \mn@doi [\mnras] {10.1093/mnras/stu1979},
  \href {http://adsabs.harvard.edu/abs/2014MNRAS.445.3939R} {445, 3939}

\bibitem[\protect\citeauthoryear{{Reiter}, {Smith}, {Kiminki}, {Bally}  \&
  {Anderson}}{{Reiter} et~al.}{2015a}]{reiter2015a}
{Reiter} M.,  {Smith} N.,  {Kiminki} M.~M.,  {Bally} J.,   {Anderson} J.,
  2015a, \mn@doi [\mnras] {10.1093/mnras/stv177}, \href
  {http://adsabs.harvard.edu/abs/2015MNRAS.448.3429R} {448, 3429}

\bibitem[\protect\citeauthoryear{{Reiter}, {Smith}, {Kiminki}  \&
  {Bally}}{{Reiter} et~al.}{2015b}]{reiter2015b}
{Reiter} M.,  {Smith} N.,  {Kiminki} M.~M.,   {Bally} J.,  2015b, \mn@doi
  [\mnras] {10.1093/mnras/stv634}, \href
  {http://adsabs.harvard.edu/abs/2015MNRAS.450..564R} {450, 564}

\bibitem[\protect\citeauthoryear{{Seward}, {Butt}, {Karovska}, {Prestwich},
  {Schlegel}  \& {Corcoran}}{{Seward} et~al.}{2001}]{seward2001}
{Seward} F.~D.,  {Butt} Y.~M.,  {Karovska} M.,  {Prestwich} A.,  {Schlegel}
  E.~M.,   {Corcoran} M.,  2001, \mn@doi [\apj] {10.1086/320961}, \href
  {http://adsabs.harvard.edu/abs/2001ApJ...553..832S} {553, 832}

\bibitem[\protect\citeauthoryear{{Shaviv}}{{Shaviv}}{2000}]{shaviv2000}
{Shaviv} N.~J.,  2000, \mn@doi [\apjl] {10.1086/312585}, \href
  {http://adsabs.harvard.edu/abs/2000ApJ...532L.137S} {532, L137}

\bibitem[\protect\citeauthoryear{{Smith}}{{Smith}}{2005}]{smith2005a}
{Smith} N.,  2005, \mn@doi [\mnras] {10.1111/j.1365-2966.2005.08750.x}, \href
  {http://adsabs.harvard.edu/abs/2005MNRAS.357.1330S} {357, 1330}

\bibitem[\protect\citeauthoryear{{Smith}}{{Smith}}{2006a}]{smith2006a}
{Smith} N.,  2006a, \mn@doi [\mnras] {10.1111/j.1365-2966.2006.10007.x}, \href
  {http://adsabs.harvard.edu/abs/2006MNRAS.367..763S} {367, 763}

\bibitem[\protect\citeauthoryear{{Smith}}{{Smith}}{2006b}]{smith2006b}
{Smith} N.,  2006b, \mn@doi [\apj] {10.1086/503766}, \href
  {http://adsabs.harvard.edu/abs/2006ApJ...644.1151S} {644, 1151}

\bibitem[\protect\citeauthoryear{{Smith}}{{Smith}}{2008}]{smith2008}
{Smith} N.,  2008, \mn@doi [\nat] {10.1038/nature07269}, \href
  {http://adsabs.harvard.edu/abs/2008Natur.455..201S} {455, 201}

\bibitem[\protect\citeauthoryear{{Smith}}{{Smith}}{2011}]{smith2011}
{Smith} N.,  2011, \mn@doi [\mnras] {10.1111/j.1365-2966.2011.18607.x}, \href
  {http://adsabs.harvard.edu/abs/2011MNRAS.415.2020S} {415, 2020}

\bibitem[\protect\citeauthoryear{{Smith} \& {Frew}}{{Smith} \&
  {Frew}}{2011}]{smithfrew2011}
{Smith} N.,  {Frew} D.~J.,  2011, \mn@doi [\mnras]
  {10.1111/j.1365-2966.2011.18993.x}, \href
  {http://adsabs.harvard.edu/abs/2011MNRAS.415.2009S} {415, 2009}

\bibitem[\protect\citeauthoryear{{Smith} \& {Gehrz}}{{Smith} \&
  {Gehrz}}{1998}]{smithgehrz1998}
{Smith} N.,  {Gehrz} R.~D.,  1998, \mn@doi [\aj] {10.1086/300447}, \href
  {http://adsabs.harvard.edu/abs/1998AJ....116..823S} {116, 823}

\bibitem[\protect\citeauthoryear{{Smith} \& {Hartigan}}{{Smith} \&
  {Hartigan}}{2006}]{smithhartigan2006}
{Smith} N.,  {Hartigan} P.,  2006, \mn@doi [\apj] {10.1086/498860}, \href
  {http://adsabs.harvard.edu/abs/2006ApJ...638.1045S} {638, 1045}

\bibitem[\protect\citeauthoryear{{Smith} \& {Morse}}{{Smith} \&
  {Morse}}{2004}]{smithmorse2004}
{Smith} N.,  {Morse} J.~A.,  2004, \mn@doi [\apj] {10.1086/382671}, \href
  {http://adsabs.harvard.edu/abs/2004ApJ...605..854S} {605, 854}

\bibitem[\protect\citeauthoryear{{Smith} \& {Tombleson}}{{Smith} \&
  {Tombleson}}{2015}]{smithtombleson2015}
{Smith} N.,  {Tombleson} R.,  2015, \mn@doi [\mnras] {10.1093/mnras/stu2430},
  \href {http://adsabs.harvard.edu/abs/2015MNRAS.447..598S} {447, 598}

\bibitem[\protect\citeauthoryear{{Smith}, {Gehrz}, {Hinz}, {Hoffmann}, {Hora},
  {Mamajek}  \& {Meyer}}{{Smith} et~al.}{2003}]{smith2003}
{Smith} N.,  {Gehrz} R.~D.,  {Hinz} P.~M.,  {Hoffmann} W.~F.,  {Hora} J.~L.,
  {Mamajek} E.~E.,   {Meyer} M.~R.,  2003, \mn@doi [\aj] {10.1086/346278},
  \href {http://adsabs.harvard.edu/abs/2003AJ....125.1458S} {125, 1458}

\bibitem[\protect\citeauthoryear{{Smith}, {Morse}  \& {Bally}}{{Smith}
  et~al.}{2005}]{smith2005b}
{Smith} N.,  {Morse} J.~A.,   {Bally} J.,  2005, \mn@doi [\aj]
  {10.1086/444562}, \href {http://adsabs.harvard.edu/abs/2005AJ....130.1778S}
  {130, 1778}

\bibitem[\protect\citeauthoryear{{Smith}, {Bally}  \& {Walborn}}{{Smith}
  et~al.}{2010}]{smith2010a}
{Smith} N.,  {Bally} J.,   {Walborn} N.~R.,  2010, \mn@doi [\mnras]
  {10.1111/j.1365-2966.2010.16520.x}, \href
  {http://adsabs.harvard.edu/abs/2010MNRAS.405.1153S} {405, 1153}

\bibitem[\protect\citeauthoryear{{Sohn}, {Anderson}  \& {van der Marel}}{{Sohn}
  et~al.}{2012}]{sohn2012}
{Sohn} S.~T.,  {Anderson} J.,   {van der Marel} R.~P.,  2012, \mn@doi [\apj]
  {10.1088/0004-637X/753/1/7}, \href
  {http://adsabs.harvard.edu/abs/2012ApJ...753....7S} {753, 7}

\bibitem[\protect\citeauthoryear{{Soker}}{{Soker}}{2007}]{soker2007}
{Soker} N.,  2007, \mn@doi [\apj] {10.1086/513711}, \href
  {http://adsabs.harvard.edu/abs/2007ApJ...661..490S} {661, 490}

\bibitem[\protect\citeauthoryear{{Stothers} \& {Chin}}{{Stothers} \&
  {Chin}}{1993}]{stotherschin1993}
{Stothers} R.~B.,  {Chin} C.-W.,  1993, \mn@doi [\apjl] {10.1086/186837}, \href
  {http://adsabs.harvard.edu/abs/1993ApJ...408L..85S} {408, L85}

\bibitem[\protect\citeauthoryear{{Thackeray}}{{Thackeray}}{1950}]{thackeray1950}
{Thackeray} A.~D.,  1950, \mnras, \href
  {http://adsabs.harvard.edu/abs/1950MNRAS.110..524T} {110, 524}

\bibitem[\protect\citeauthoryear{{Walborn}}{{Walborn}}{1976}]{walborn1976}
{Walborn} N.~R.,  1976, \mn@doi [\apjl] {10.1086/182045}, \href
  {http://adsabs.harvard.edu/abs/1976ApJ...204L..17W} {204, L17}

\bibitem[\protect\citeauthoryear{{Walborn} \& {Blanco}}{{Walborn} \&
  {Blanco}}{1988}]{walbornblanco1988}
{Walborn} N.~R.,  {Blanco} B.~M.,  1988, \mn@doi [\pasp] {10.1086/132237},
  \href {http://adsabs.harvard.edu/abs/1988PASP..100..797W} {100, 797}

\bibitem[\protect\citeauthoryear{{Walborn}, {Blanco}  \& {Thackeray}}{{Walborn}
  et~al.}{1978}]{walborn1978}
{Walborn} N.~R.,  {Blanco} B.~M.,   {Thackeray} A.~D.,  1978, \mn@doi [\apj]
  {10.1086/155806}, \href {http://adsabs.harvard.edu/abs/1978ApJ...219..498W}
  {219, 498}

\bibitem[\protect\citeauthoryear{{Weis}}{{Weis}}{2012}]{weis2012}
{Weis} K.,  2012, in {Davidson} K.,  {Humphreys} R.~M.,  eds,  Astrophysics and
  Space Science Library Vol. 384, Astrophysics and Space Science Library.
  p.~171, \mn@doi{10.1007/978-1-4614-2275-4_8}

\bibitem[\protect\citeauthoryear{{Weis}, {Duschl}  \& {Bomans}}{{Weis}
  et~al.}{2001}]{weis2001}
{Weis} K.,  {Duschl} W.~J.,   {Bomans} D.~J.,  2001, \mn@doi [\aap]
  {10.1051/0004-6361:20000460}, \href
  {http://adsabs.harvard.edu/abs/2001A%26A...367..566W} {367, 566}

\bibitem[\protect\citeauthoryear{{Whitelock}, {Feast}, {Marang}  \&
  {Breedt}}{{Whitelock} et~al.}{2004}]{whitelock2004}
{Whitelock} P.~A.,  {Feast} M.~W.,  {Marang} F.,   {Breedt} E.,  2004, \mn@doi
  [\mnras] {10.1111/j.1365-2966.2004.07950.x}, \href
  {http://adsabs.harvard.edu/abs/2004MNRAS.352..447W} {352, 447}

\bibitem[\protect\citeauthoryear{{de Jager}}{{de Jager}}{1984}]{dejager1984}
{de Jager} C.,  1984, \aap, \href
  {http://adsabs.harvard.edu/abs/1984A%26A...138..246D} {138, 246}

\bibitem[\protect\citeauthoryear{{van Genderen}}{{van
  Genderen}}{2001}]{vangenderen2001}
{van Genderen} A.~M.,  2001, \mn@doi [\aap] {10.1051/0004-6361:20000022}, \href
  {http://adsabs.harvard.edu/abs/2001A%26A...366..508V} {366, 508}

\makeatother
\end{thebibliography}

%%%%%%%%%%%%%%%%%%%%%%%%%%%%%%%%%%%%%%%%%%%%%%%%%%%%%%%%%%%%%%%%%%%%%%%

\bsp	
\label{lastpage}
\end{document}